\documentclass{emulateapj}

\begin{document}

\title{Photometric properties of Ly$\alpha$ emitters at $z \approx 4.86$ in the COSMOS 2 square degree field\altaffilmark{1}}

\author{Y. Shioya,\altaffilmark{2} 
        Y. Taniguchi,\altaffilmark{2} 
        S. S. Sasaki,\altaffilmark{3, 4} 
        T. Nagao,\altaffilmark{2, 3} 
        T. Murayama,\altaffilmark{4} 
        T. Saito,\altaffilmark{2} 
        Y. Ideue,\altaffilmark{3} 
        A. Nakajima,\altaffilmark{3} 
        K. L. Matsuoka,\altaffilmark{3} 
        J. Trump,\altaffilmark{2,5}
        N. Z. Scoville,\altaffilmark{6} 
	D. B. Sanders,\altaffilmark{7} 
	B. Mobasher,\altaffilmark{8} 
	H. Aussel,\altaffilmark{9} 
        P. Capak,\altaffilmark{6,10} 
        J. Kartaltepe,\altaffilmark{7} 
	A. Koekemoer,\altaffilmark{11} 
        C. Carilli,\altaffilmark{12} 
        R. S. Ellis,\altaffilmark{6, 13} 
        B. Garilli,\altaffilmark{14} 
        M. Giavalisco,\altaffilmark{11} 
        M. G. Kitzbichler,\altaffilmark{15} 
        C. Impey,\altaffilmark{5}
        O. LeFevre,\altaffilmark{16} 
        E. Schinnerer,\altaffilmark{17} and 
        V. Smolcic\altaffilmark{18, 19}
		}

\altaffiltext{1}{Based on data collected at 
	Subaru Telescope, which is operated by 
	the National Astronomical Observatory of Japan.}
\altaffiltext{2}{Research Center for Space and Cosmic Evolution, Ehime University, 
        Bunkyo-cho, Matsuyama 790-8577, Japan}
\altaffiltext{3}{Graduate School of Science and Engineering, Ehime University, 
        Bunkyo-cho, Matsuyama 790-8577, Japan}
\altaffiltext{4}{Astronomical Institute, Graduate School of Science,
        Tohoku University, Aramaki, Aoba, Sendai 980-8578, Japan}
\altaffiltext{5}{Steward Observatory, University of Arizona, Tucson, AZ 85721}
\altaffiltext{6}{Department of Astronomy, MS 105-24, California Institute of
                Technology, Pasadena, CA 91125}
\altaffiltext{7}{Institute for Astronomy,  University of Hawaii,
                 2680 Woodlawn Drive, HI 96822}
\altaffiltext{8}{Department of Physics and Astronomy, 
                 University of California, Riverside, CA, 92521}
\altaffiltext{9}{AIM Unit\'e Mixte de Recherche CEA CNRE Universit\'e Paris VII UMR n158 France}
\altaffiltext{10}{Spitzer Science Center, California Institute of
                Technology, Pasadena, CA 91125}
\altaffiltext{11}{Space Telescope Science Institute, 3700 San Martin Drive,
                 Baltimore, MD 21218}
\altaffiltext{12}{National Radio Astronomy Observatory,
                  P.O. Box 0, Socorro, NM 87801-0387}
\altaffiltext{13}{Department of Astrophysics Keble Road, Oxford OX2 3RH, UK}
\altaffiltext{14}{INAF, Istituto di Astrofisica Spaziale e Fisica Cosmica,
                  Sezione di Milano, via Bassini 15, 20133 Milano, Itary}
\altaffiltext{15}{Max-Planck-Institut f\"ur Astrophysik,
                  D-85748 Garching bei M\"unchen, Germany}
\altaffiltext{16}{Laboratoire d'Astrophysique de Marseille,
                  BP 8, Traverse du Siphon, 13376 Marseille Cedex 12, France}
\altaffiltext{17}{Max Planck Institut f\"ur Astronomie,
                  K\"onigstuhl 17, Heidelberg, D-69117, Germany}
\altaffiltext{18}{Princeton University Observatory, Princeton, NJ 08544}
\altaffiltext{19}{University of Zagreb, Department of Physics, Bijenicka cesta 32, 
                  10000 Zagreb, Croatia}

\shortauthors{Shioya et al.}
\shorttitle{Ly$\alpha$ emitters at $z = 4.86$}

\begin{abstract}
We present results of a survey for Ly$\alpha$ emitters at $z \approx 4.86$ based on 
optical narrowband ($\lambda_c=7126$\AA, $\Delta \lambda=73$\AA) and 
broadband ($B$, $V$, $r^\prime$, $i^\prime$, and $z^\prime$) observations of 
the Cosmic Evolution Survey (COSMOS) field using Suprime-Cam on the Subaru Telescope. 
We find 79 LAE candidates at $z\approx 4.86$ over a contiguous survey area of 1.83 deg$^2$, 
down to the Ly$\alpha$ line flux of $1.47 \times 10^{-17}{\rm ergs \; s^{-1} \; cm^{-2}}$.
We obtain the Ly$\alpha$ luminosity function with a best-fit Schechter parameters 
of $\log L^*=42.9^{+0.5}_{-0.3}{\rm ergs \; s^{-1}}$ and 
$\phi^* = 1.2^{+8.0}_{-1.1} \times 10^{-4}{\rm Mpc^{-3}}$ for $\alpha=-1.5$ (fixed). 
The two-point correlation function for our LAE sample is 
$\xi(r) = (r/4.4^{+5.7}_{-2.9}{\rm Mpc})^{-1.90\pm0.22}$. 

In order to investigate the field-to-field variations of the properties of Ly$\alpha$ 
emitters, we divide the survey area into nine tiles of $0.5^\circ\times 0.5^\circ$ each. 
We find that the number density varies with a factor of $\simeq 2$ from field to field with high statistical significance. 
However, we find no significant field-to-field variance when we divide the field into 
four tiles with $0.7^\circ\times 0.7^\circ$ each. We conclude that at least 0.5 deg$^2$ survey area is required 
to derive averaged properties of LAEs at $z\sim 5$, and our survey field is wide enough 
to overcome the cosmic variance. 

\end{abstract}


\section{INTRODUCTION}

Study of the formation and evolution of galaxies is among the most important 
topics in modern astrophysics. An essential component of such investigations 
is the identification of galaxies at the highest redshifts, 
when most of the galaxies formed, and study of their rest-frame properties. 
This requires multi-waveband, wide-area and deep surveys of galaxies to
provide statistically significant population of these objects. Recently, 
complementary observations of selected fields by the largest ground-based 
and space-borne telescopes have made this aim possible by extending this study
to $z\sim 7$ and providing statistically large samples of high redshift 
galaxies with a significant fraction of them confirmed spectroscopically
(see Taniguchi 2008 for a recent review). 

To summarize, there are two established techniques to search for high-$z$ 
galaxy candidates. Firstly, the Lyman break method (i.e. also called 
drop-out technique) which identifies the
continuum break characteristic of Lyman alpha absorption by the Inter-Galactic
Medium (IGM) (Steidel et al 1996; for a review see Giavalisco 2002). The
high-$z$ candidates selected by this technique are called Lyman break galaxies 
(LBGs). Secondly, the narrow-band imaging technique which aims for detection of
galaxies with strong Lyman $\alpha$ emission- so called  
Lyman alpha emitters (LAEs) (Taniguchi et al. 2003 for a review).  
Although both LBGs and LAEs are actively star-forming galaxies, there are
systematic differences between them. For example, 
the stellar populations of LAEs are relatively younger, 
they have a smaller stellar mass (e.g., Lai et al. 2008), smaller size 
(e.g., Dow-Hygelund et al. 2007) and are less dusty (e,g, Shapley et al. 2003) compared
to the LBGs. These observations imply that the LAEs are likely to be 
in an earlier star formation phase with respect to LBGs. Furthermore, it is
estimated that the average mass of dark matter halos hosting LAEs and LBGs at 
$z \sim 4$ -- 5 ($\sim 10^{12}M_\odot$) are comparable 
(Ouchi et al. 2004; Kova{\v c} et al. 2007), while,  
at $z \sim 3.1$ ($\sim 10^{11}M_\odot$), it is smaller for the LAEs
(Gawiser et al. 2007). 
This implies that the LAEs at $z \sim 3.1$ are likely progenitors of 
present-day $L^*$ galaxies, whereas the LAEs at $z \sim 4$ -- 5 
and LBGs at $z \sim 3$ -- 5 will evolve 
into present-day galaxies with $L>2.5L^*$(Gawiser et al. 2007). 

To understand differences between the LAEs and LBGs
at any given redshift and their properties with look-back time, 
one needs statistically
large and complete samples of these galaxies at different redshifts. 
Specially for the LAEs, due to technical difficulties in 
performing narrow-band observations, the majority of 
these surveys are performed over small areas and in selected redshift slices
where there are windows to avoid absorption of the lines by the atmosphere.   
This problem is particularly serious for candidates at higher redshifts
where one needs both depth and wide-area coverage to have sufficient number
of galaxies and to minimize the cosmic variance. 

For the LAEs at $z \sim 5.7$, extensive studies in different fields
have been performed, including: survey around quasar SDSS J1044-0125 
(Ajiki et al. 2003), SSA22 (Hu et al. 2004), 
GOODS-N and GOODS-S (Ajiki et al. 2006), 
the Subaru Deep Field (SDF) (Shimasaku et al. 2006), 
the Subaru-XMM Newton Deep Field (SXDF) (Ouchi et al. 2005, 2008) and 
the Cosmic Evolution Survey (COSMOS) (Murayama et al. 2007). However, 
there are only limited surveys of LAEs at other redshifts. This is a
serious deficiency in studying evolution of clustering of the LAEs and
their rest-frame properties specially if these are expected to evolve to 
nearby elliptical galaxies (Gawiser et al. 2007).

In this paper we perform the largest survey of the LAEs at $z \approx 4.86$, 
covering the entire 2 square degree of the COSMOS field 
(Scoville et al. 2007). Earlier studies of the LAEs at this redshift revealed
presence of large scale structures of $20 \times 50$ Mpc size 
(Ouchi et al. 2003; Shimasaku et al. 2003), that are comparable to almost the size of the
surveyed area, indicating  serious cosmic variance in these data
(Shimasaku et al. 2003). The survey performed in this study covers an
area of 190 Mpc $\times$ 190 Mpc [7 times larger than 
the survey area of Shimasaku et al. (2003, 2004)], large enough to 
encompass structures of $\sim 50\times 50$ Mpc$^2$ size, allowing for 
proper sampling of the average properties of LAEs at $z \sim 4.9$. 
Therefore, we are able to examine how the cosmic variance affects the derivation of 
both the Ly$\alpha$ luminosity functions and the clustering properties for 
the first time. 

In the next section we present our sample selection of LAEs. 
In section 3 we discuss the Ly$\alpha$ luminosity function and 
the clustering properties of our sample. 
We summarize our results in section 4. 
Throughout this paper, magnitudes are given in the AB system. 
We adopt a flat universe with $\Omega_{\rm matter}=0.3$, $\Omega_\Lambda=0.7$, 
and $H_0=70 \; {\rm km \; s^{-1} \; Mpc^{-1}}$. 

\section{THE SAMPLE}

\subsection{The Data}

We carried out an optical narrow-band (NB711; $\lambda_c=7126$\AA, 
$\Delta \lambda=73$ \AA) imaging survey of the entire 
2-deg$^2$ area of the COSMOS field, using the Suprime-Cam on 
the Subaru Telescope. 
The NB711 observations were done on February 2006 (Taniguchi et al. 2008). 
The data were reduced using the IMCAT software.\footnote{IMCAT is distributed by Nick Kaiser at http://www.ifa.hawaii.edu/~kaiser/imcat/ .} 
Combining the NB711 data with the broad-band ($B$, $V$, $r^\prime$, 
$i^\prime$, and $z^\prime$) Suprime-Cam imaging data and $i^*$-band 
Mega-Prime/CFHT imaging data already available 
(Taniguchi et al. 2007; Capek et al. 2007),\footnote{http://irsa.ipac.caltech.edu/data/COSMOS/} 
we identified LAE candidates at $z \approx 4.86$.  
Details of the narrow-band and broad-band observations and 
data reduction are presented by Taniguchi et al. (2007, 2008) and 
Capak et al. (2007). 

The FWHMs corresponding to the PSFs on the final images are 0{\farcs}95 ($B$), 1{\farcs}33 ($V$),
1{\farcs}05 ($r^\prime$), 0{\farcs}95 ($i^\prime$), 1{\farcs}15 ($z^\prime$) and 
0{\farcs}79 ({\it NB711}). The images were all degraded to a PSF size of 1{\farcs}33.
 The limiting AB magnitudes of the final PSF matched images are: 
$B=27.56$, $V=26.77$, $r^\prime=26.95$, $i^\prime=26.49$, $z^\prime=25.45$, 
and $\mathit{NB711}=25.17$ for a $3\sigma$ detection in a 
$3^{\prime \prime}$ diameter aperture. 
We then performed source detection on the original {\it NB711} image 
using SExtractor (Bertin \& Arnouts 1996), followed by photometry
over an aperture of $3^{\prime \prime}$ diameter as described in
Capak et al (2007). Similarly, $i^*$ band (CFHT) magnitudes over the same
aperture, are used to identify interlopers consisting of bright galaxies 
with $i^\prime < 22$. 

After subtracting the masked-out regions, the effective survey area 
is 1.83 $\rm deg^2$. 
The redshift coverage of NB711 is $4.83 \le z \le 4.89$ ($\Delta z = 0.06$), 
giving an effective survey volume of $1.1 \times 10^6 \; {\rm Mpc^3}$ (comoving).

\subsection{Selection of Lyman $\alpha$ Emitters at $z \approx 4.86$}

In order to select NB711-excess objects efficiently, we first 
need the magnitude of a frequency-matched continuum. 
Since the effective frequency of the NB711 filter (421.1 THz) lies between 
$r^{\prime}$ (482.8 THz) and $i^\prime$ (394.0 THz) bands, we estimate the 
frequency-matched continuum, ``$ri$ continuum'', using the 
linear combination: $f_{ri} = 0.3 f_{r^\prime}+0.7 f_{i^\prime}$, 
where $f_{r^\prime}$ and $f_{i^\prime}$ are the 
flux densities in $r^{\prime}$  and $i^\prime$ bands respectively. This gives
a 3 $\sigma$ limiting magnitude of $ri_{\rm lim,3\sigma}=26.84$ for the continuum 
(in a $3^{\prime \prime}$ diameter aperture). 
Since brighter objects with $i^\prime < 22$ are saturated in Subaru images, 
we use the $i^*$ flux density, $f_{i^*}$, to calculate the ``{\it ri} continuum'' for such 
objects, i.e., $f_{ri} = 0.3 f_{r^\prime}+0.7 f_{i^*}$. 

The {\it NB}711-excess objects are then selected using the following criteria: 
\begin{eqnarray}
ri-NB711 & > & 0.7, \; \; \; {\rm and}\\
ri-NB711 & > & 3 \sigma(ri-NB711),
\end{eqnarray}
where $3\sigma(ri-NB711)=-2.5\log
\left( 1-\sqrt{(f_{3\sigma_{\mathit{NB}711}})^2+(f_{3\sigma_{ri}})^2}/f_{\mathit{NB}711} \right)$.

\begin{figure}
\epsscale{0.5}
\plotone{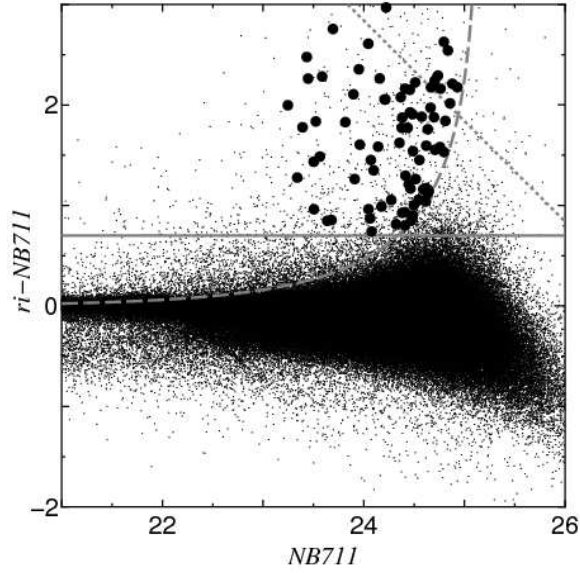}
\caption{
$ri-NB711$ vs. $NB711$ diagram for all the NB711 detected sources in COSMOS.  
The horizontal solid line corresponds to $ri-NB711=0.7$. 
The dashed line show the distribution of the $3\sigma$ errors. 
The dotted line shows the limiting magnitude of $ri$. 
Filled circles represent 79 LAE candidates detected here. 
\label{La:cm}
}
\end{figure}

The first criterion corresponding to an observed equivalent width of $EW_{\rm obs} > 66$ \AA, 
means the flux density in narrow band is twice as large as the flux density of {\it ri} continuum. 
This kind of criterion is conventionally used for LAE survey (e.g., Ouchi et al. 2003; Ajiki et al. 
2006; Murayama et al. 2007) to select reliable emitter candidates. 
Taking account the scatter of the $ri-NB711$ color, 
we added the second criterion. 
These two criteria are shown in Figure \ref{La:cm}. 
For objects with $ri < ri_{\rm lim,3\sigma}$, 
we use the lower-limit value of the {\it NB711}-excess, 
$(ri-NB711)_{\rm low.limit}=ri_{\rm lim,3\sigma}-NB711$, 
for our sample selection.  
We finally select the NB711-excess objects with $NB711<24.9$. 

\begin{figure}
\epsscale{0.6}
\plotone{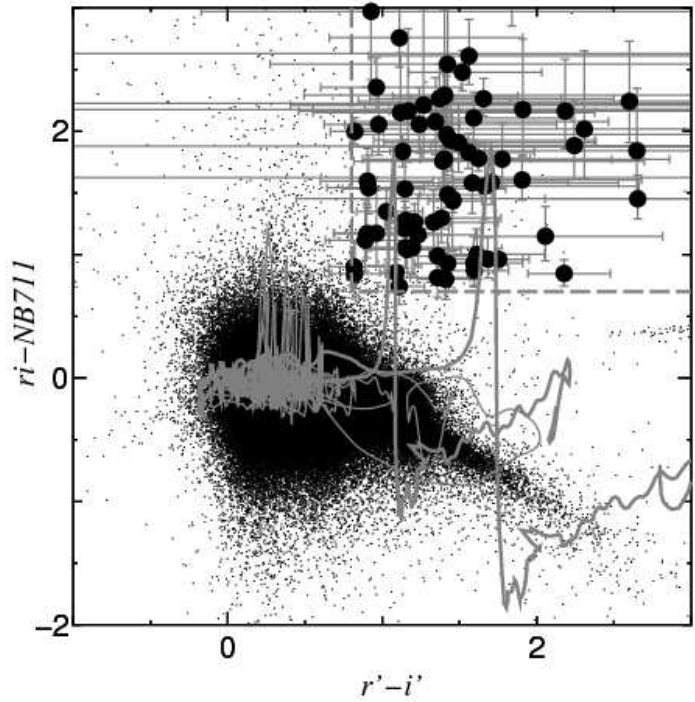}
\caption{
$ri-NB711$ vs. $r^\prime-i^\prime$ diagram. 
Thick gray lines show colors of model LAE SEDs, which are calculated with BC03 model 
(Bruzual \& Charlot 2003) with an exponential decay time of tau=1Gyr and an age of 1Gyr, 
corresponding to a cosmic transmission of 
$0.5 \tau_{\rm eff}$ (left) and $\tau_{\rm eff}$ (right) in the formulation of Madau et al. (1996). 
Luminosity of the Ly$\alpha$ emission is calculated as 
$L({\rm Ly}\alpha) = 1.2 \times 10^{-11} N_{\rm Lyc}$, 
where $N_{\rm Lyc}$ is the ionizing photon production rate 
(Leitherer \& Heckman 1995; Brocklhurst 1971). 
Thin gray lines show color loci of starburst galaxies (Kinney et al. 1996), 
typical elliptical, spiral, and irregular galaxies (Coleman, Wu, \& Weedman 1980) 
up to $z=2$. 
\label{La:cc}
}
\end{figure}

Following the above criteria, we find a total of 1154 NB711-excess objects.
These objects includes not only LAEs at $z \approx 4.86$, but other low-$z$ interlopers 
such as H$\alpha$, [O{\sc iii}], and [O{\sc ii}] emitting galaxies. 
In order to distinguish LAEs from low-$z$ interlopers, we compare
the observed broad-band colors of the LAE candidates with colors 
that are estimated by using the model spectral energy distribution derived by Coleman, Wu, 
\& Weedman (1980), Kinney et al. (1996), and Bruzual \& Charlot (2003). 
Figure \ref{La:cc} shows the $ri-NB711$ vs. $r^\prime-i^\prime$ color-color diagram
with the predicted colors overlaid. 
Because of the cosmic transmission (Madau et al. 1996), 
the $r^\prime-i^\prime$ colors of LAEs 
are predicted to be redder than low-$z$ emission-line galaxies. 
Based on results from Figure \ref{La:cc}, we added another condition to the selection criteria:
\begin{equation}
r^\prime-i^\prime > 0.8 \; .
\end{equation}

\begin{figure}
\epsscale{0.4}
\plotone{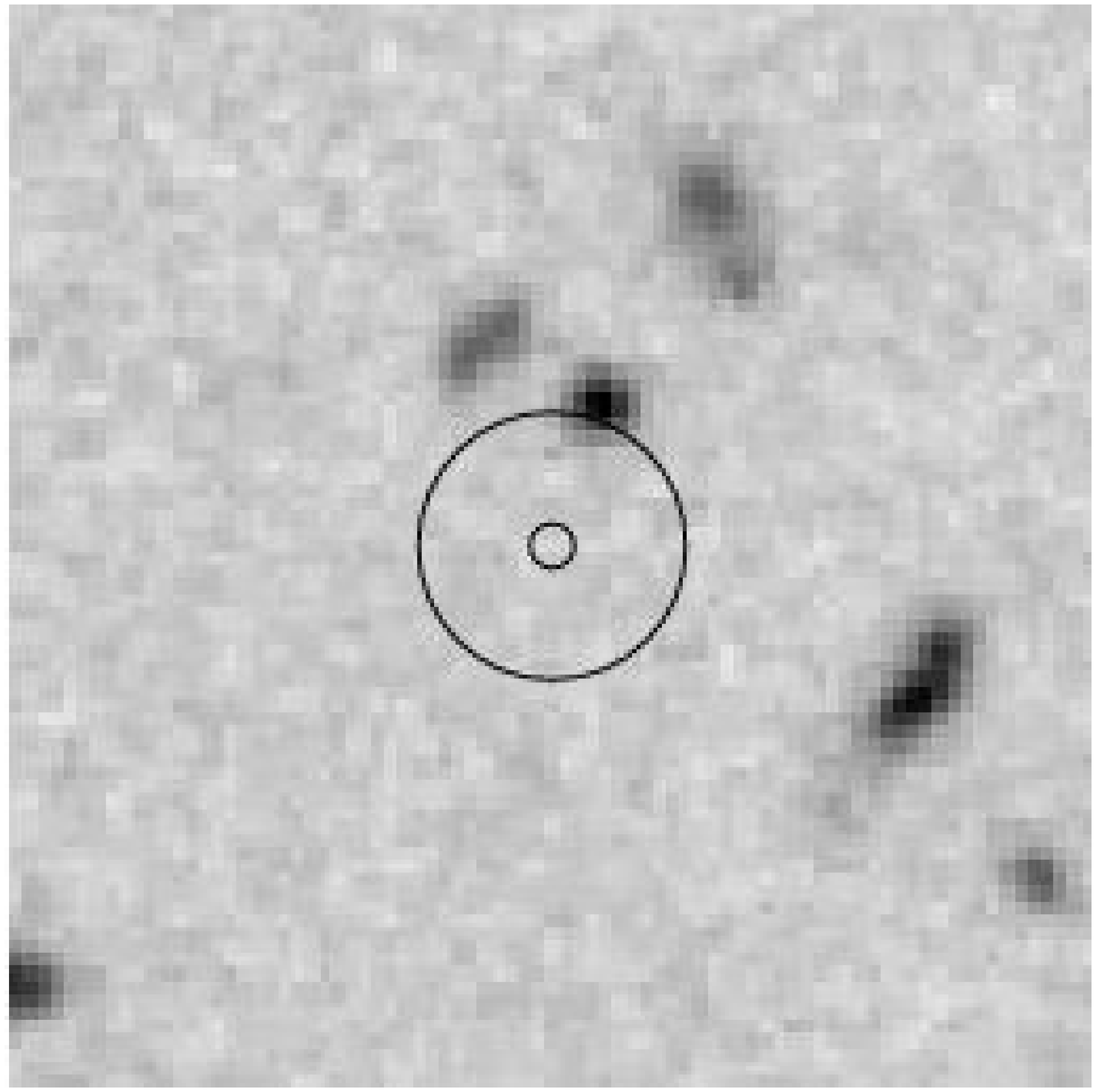}
\plotone{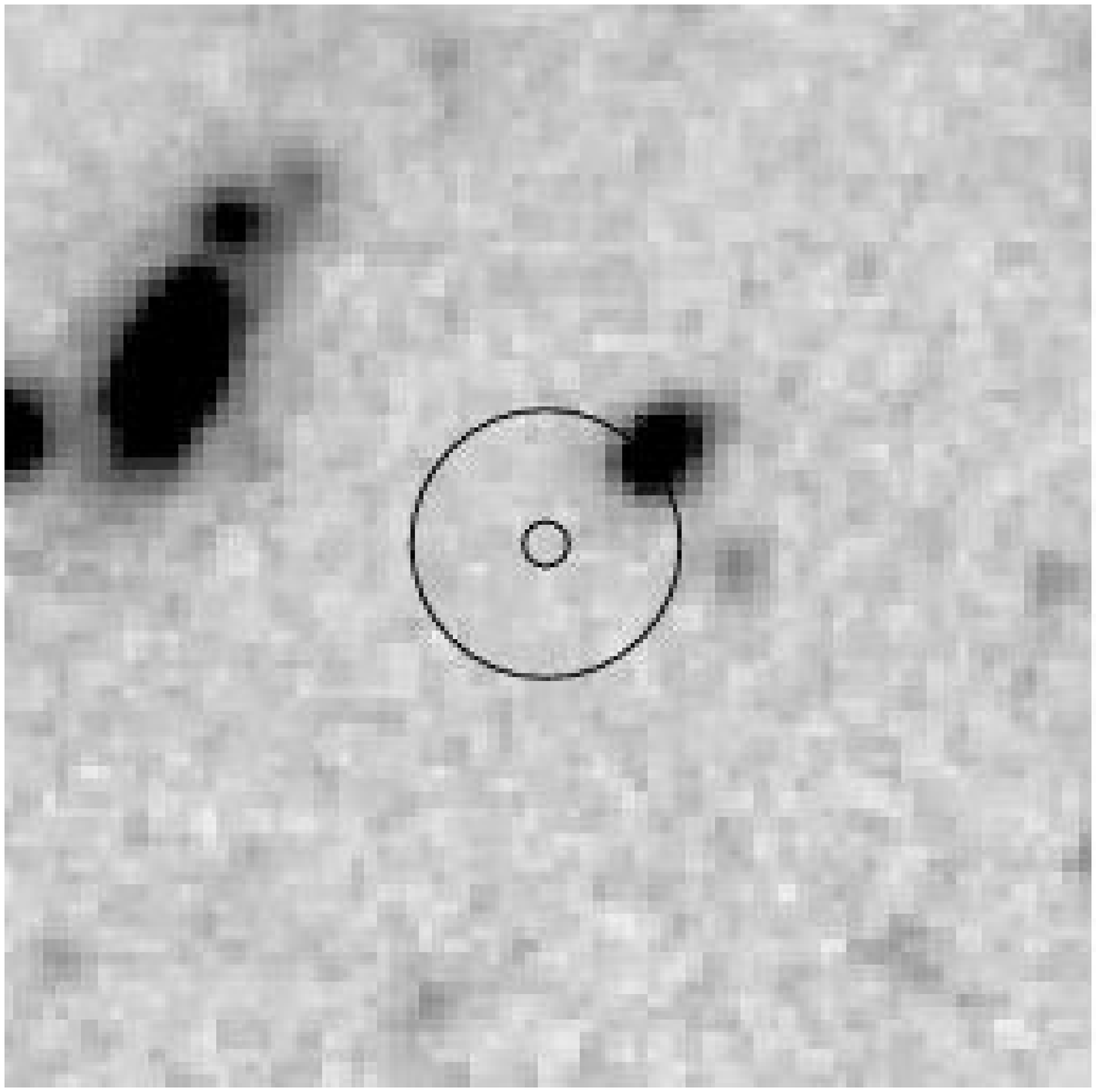}
\caption{
$B$-band images of our LAE candidates \#16 \& \#44. 
Each box is $12^{\prime \prime}$ on a side (north is up and east is left). 
The diameter of a small (large) circle is $0{\farcs}5$ ($3^{\prime \prime}$). 
In both cases, there are no counter part in the center, 
although the flux within $3^{\prime \prime}$ diameter aperture is larger than 
$3\sigma$ because of the contamination from nearby objects. 
\label{La:thumB}
}
\end{figure}

Since the Lyman break is redshifted to $\sim 5300$ \AA, the $B$-band flux of LAEs at 
$z \approx 4.86$ is expected to be zero. 
The $B$-dropout is an effective criterion to distinguish LAEs from low-$z$ interlopers. 
Here, we must pay attention to the contamination from nearby objects on the sky. 
If there are objects detected in $B$-band near the LAE and the fluxes from these objects 
in the aperture are not negligible, the LAE may be misclassified as a low-$z$ interloper. 
We show $B$-band images of two of our final LAE candidates in Figure \ref{La:thumB}. 
Although there is no object at the position of the emission-line object (center), 
the $B$-band magnitude in $3^{\prime \prime}$ diameter aperture is brighter than 27.56 ($3\sigma$), 
because of the contamination from the object that lie at a distance of $\sim 1.5^{\prime \prime}$ 
from the center. 
To avoid the contamination from such nearby objects, 
we adopt the $B$-band magnitude measured with in the small aperture ($0{\farcs}5 \phi$) in the original image.
We therefore add another selection criterion, 
\begin{equation}
B_{\rm original}(0{\farcs}5 \phi) > 30.09 \; ,
\end{equation}
where $B_{\rm original}(0{\farcs}5 \phi)$ is the $B$-band magnitude over a $0{\farcs}5$ 
diameter aperture, measured in the original image (i.e. the image before
convolving and with a PSF size of $0{\farcs}95$). 
The $3\sigma$ limiting magnitude for a $0{\farcs}5 \phi)$ aperture in the original image 
is 30.09. 

Based on the added criteria, we can now clearly distinguish LAEs from the low-$z$
interlopers. We finally select a total of 79 LAE candidates at $z\sim 4.86$. 
The photometric properties of these LAE candidates are 
listed in Table 1. Their broad- and {\it NB711}-band images are presented in Figure 3.

To further check the validity of our photometric selection and
their expected redshifts, we extract information of the LAE candidates from
the COSMOS spectroscopic catalogue. 
A total of 7 LAEs in our final candidate list 
have spectroscopic redshifts. Figure 4 presents the spectroscopic redshift 
distribution of these LAEs. This peaks at $z \approx 4.85$ with all 
the spectroscopic redshifts lying in the range $4.80 < z < 4.85$. 
This result suggests that our selection criteria works quite well to identify 
LAEs at $z \approx 4.9$. 

\subsection{Ly$\alpha$ Luminosity}

We estimate the line fluxes for our LAE candidates, $F_{\rm L}$, using the 
prescription by Pascual et al. (2001). 
We express the flux density in each filter band using the line flux, 
$F_L$, and the continuum flux density, $f_C$: 
\begin{eqnarray}
f_{\rm NB} & = & f_C + \frac{F_L}{\rm \Delta NB} \label{eqn:fnb}\\
f_{r^\prime} & = & f_C \\
f_{i^\prime} & = & f_C + \frac{F_L}{\Delta i^\prime},
\end{eqnarray}
where $\rm \Delta NB$ and $\Delta i^\prime$ are the effective bandwidths of 
NB711 and $i^\prime$, respectively. 
The $ri$ continuum is expressed as 
\begin{equation}
f_{ri} = 0.3 f_{r^\prime} + 0.7 f_{i^\prime} = f_C + 0.7 \frac{F_L}{\Delta i^\prime}. 
\label{eqn:fri}
\end{equation}
Using equations (\ref{eqn:fnb}) and (\ref{eqn:fri}), the line flux, $F_L$, is 
calculated by 
\begin{equation}
F_{\rm L} = \Delta NB \frac{f_{\rm NB} - f_{ri}}{1-0.7(\Delta NB/\Delta i^\prime)}. 
\end{equation}
The limiting line flux of our survey is $1.47 \times 10^{-17}{\rm ergs \; s^{-1}}$. 
Since the response curve is not square in shape, the observed flux of Ly$\alpha$ emission 
for a fixed Ly$\alpha$ luminosity depends on the redshift. On average, the observed flux is 
underestimated by a factor of 0.81, which is calculated by 
\[
\frac{\int_{\lambda_c - \Delta \lambda/2}^{\lambda_c + \Delta \lambda/2} R(\lambda) d \lambda} 
{\int_{\lambda_c - \Delta \lambda/2}^{\lambda_c + \Delta \lambda/2} d \lambda}
= 0.81 \; , 
\]
where $R(\lambda)$ is a response function normalized by the maximum value. 
We therefore apply correction statistically for the filter response by 
$F_{\rm cor}({\rm Ly}\alpha) = F_{\rm L} \times 1.24$. 
Finally, we estimate the Ly$\alpha$ luminosity as
 $L({\rm Ly}\alpha) = 4\pi d_{\rm L}^2F_{\rm cor}({\rm Ly}\alpha)$.
In this procedure, we assume that all the LAEs are located at $z = 4.86$ ($d_{\rm L}=45.1$ Gpc), 
the redshift corresponding to the central wavelength of our {\it NB}711 filter. 

\section{RESULTS AND DISCUSSION}

\subsection{Spatial Distribution}
\label{sec:SD}

Figure \ref{La:radec} presents the spatial distribution of our 79 LAE candidates at 
$z\sim 4.86$. The contours of local surface density ($2\bar{\Sigma}$, where $\bar{\Sigma}$ is 
the averaged surface density over the whole field, 43 degree$^{-2}$) are shown in the figure. 
The local surface density at position ($x$, $y$) is the density averaged over the circle 
centered at ($x$, $y$), whose radius is determined as the angular distance to the 3rd 
nearest neighbors. There are ten overdensity regions in the field. 
A typical size of the large overdensity region is $0.4^\circ \times 0.2^\circ$ (50 Mpc $\times$ 25 Mpc), 
being similar to those found by Shimasaku et al. (2003) and Ouchi et al. (2005). 

To check for field-to-field variation, 
we divide the survey area into nine subfields, each corresponding to 
a sky area of $0.5^\circ \times 0.5^\circ$ (63 Mpc $\times$ 63 Mpc) (Figure \ref{La:radec}). 
The number density of the LAEs in each subfield is summarized in Table \ref{La:tab:CV}. 
We find significant field-to-field variations among the nine subfields by a factor of $\simeq 2$. 
This means that the typical scale of the large scale structure is comparable to the size of the subfield, 
that is consistent with the size of the overdensity regions found in the above. 
The field-to-field variations found here, agrees with
those for LAEs at $z \approx 5.7$, independently estimated in the SXDF 
(Ouchi et al. 2008) and with theoretical predictions using 
the cosmological hydrodynamic simulations (Nagamine et al. 2008) for the fields of $\sim 0.2 {\rm deg^2}$.
Our finding suggests that the derived properties of LAEs from the survey with a 
small survey area (smaller than $0.5^\circ \times 0.5^\circ$) may be affected 
by the cosmic variance. 

We also divide the survey area into four subfields, each corresponding to 
a sky area of $0.7^\circ \times 0.7^\circ$ (95 Mpc $\times$ 95 Mpc). 
We find 21, 21, 19 and 18 LAEs in NE, NW, SW, and SE quadrant, respectively. 
This means that the typical scale of the large scale structure is smaller than the size of the subfield. 
Our finding suggests that the derived properties of LAEs from the survey with a 
large survey area (larger than $0.7^\circ \times 0.7^\circ$) are considered to be averaged ones 
over the universe at $z \sim 5$. 

\subsection{Ly$\alpha$ Luminosity Function}

The rest-frame Ly$\alpha$ luminosity function (LF) for our sample of
LAEs at $z\approx 4.86$ is presented in 
Figure \ref{La:LaLF}. 
The LF is measured as 
\begin{equation}
\Phi(\log L_i) = \frac{N_i}{V_{\rm co}}, 
\end{equation}
where $V_{\rm co}$ is the comoving volume of $1.1 \times 10^6{\rm Mpc^3}$ ($4.83 \le z \le 4.89$) 
and $N_i$ is a number of LAEs within 
$\log L_i \pm \frac{1}{2} \Delta \log L$. 
We use $\Delta \log L({\rm Ly\alpha}) = 0.2$. 
We fit the rest-frame Ly$\alpha$ LF with the Schechter function 
(Schechter 1976) 
using parametric maximum likelihood estimator 
(Sandage, Tammann, \& Yahil 1979). Since the characteristic luminosity ($L^*$)
and the faint-end slope ($\alpha$) of the Schechter LFs are not
independent, 
we perform the fit by fixing $\alpha$ to $-1$, $-1.5$, and $-2$.   
Our best-fit Schechter parameters are summarized in Table \ref{La:tab:LALF}. 
For comparison, we also plot the Ly$\alpha$ LF for a sample selected by Ouchi et al. (2003). 
Their survey was performed by using the same narrowband filter (NB711), 
for smaller field (543 arcmin$^2$) and deeper (down to $\mathit{NB711}=26.0$) than ours. 
Although our sample does not include low-luminosity LAEs and their sample does not include 
LAEs at the luminous-end, our Ly$\alpha$ LF is consistent with that of Ouchi et al. (2003) 
for the range of $42.8 \le \log L({\rm Ly}\alpha) \le 43.2$. 

Since the filter response curve of NB711 is not box-shaped,  
the narrow-band magnitude of LAEs of a fixed Ly$\alpha$ luminosity 
varies as a function of the redshift. 
The selection function of LAEs in terms of the equivalent width also 
changes with the redshift (Shimasaku et al. 2006; Ouchi et al. 2008). 
We check the validity of the Ly$\alpha$ LF derived above. 
In order to examine whether or not we can 
reconstruct an input Ly$\alpha$ LF by our selection criteria and 
an estimation of Ly$\alpha$ flux, we performed the Monte Carlo simulations 
that are similar to those made by Shimasaku et al. (2006). 
First, we generate a mock catalog of LAEs with a set of the Schechter 
parameters ($\alpha$, $\phi^*$, $L^*$) and a Gaussian distribution function of $EW$, 
$f(EW_0)dEW_0 \propto \exp (-EW_0^2/2\sigma_{EW})dEW_0$. 
We adopt four $\sigma_{EW}$ values: $\sigma_{EW}=50$, 100, 200, and 400 \AA. 
We uniformly distribute them in comoving space over $4.7 < z < 5.1$. 
Next, we select Ly$\alpha$ emitters and evaluate the Ly$\alpha$ LF applying 
the method written above for the mock catalog. 
We show results of our simulations in Figure \ref{La:LaLFmock}. 
We confirm that the Ly$\alpha$ LF we evaluate is very close to the input LF. 
We conclude that the simple method we adopted is valid for evaluating the Ly$\alpha$ LF. 

We plot the Ly$\alpha$ LFs from the four subfields in Figure \ref{La:LaLF} ({\it left panel}). 
Those LFs are consistent within their errors. 
We also summarize the best-fit Schechter parameters for four subfields in Table \ref{La:tab:LALF2}. 
Although the field-to-field variation of $\phi^*$ is a factor of 4, 
each value exist within the error in Table \ref{La:tab:LALF}. 
In Figure \ref{La:LaLF} ({\it right panel}), we compare our results with 
other LAE surveys in the redshift range $z \sim 3.1$ -- 6.6. 
Although various surveys have slightly different selection criteria, 
most of the Ly$\alpha$ LFs are similar to each other. 
We then find that estimated Ly$\alpha$ LF is very similar to those at $3.1 \le z \le 5.7$ within errors. 
This result supports the little evolution of Ly$\alpha$ LFs in the range of $3 < z < 6$ 
(Tran et al. 2004; van Breukelen et al. 2005; Shimasaku et al. 2006; Ouchi et al. 2008). 

\subsection{Equivalent Widths}

Figure \ref{La:ewdf} shows the distribution of $EW_0({\rm Ly}\alpha)$. 
To measure the rest-frame UV continuum flux, we use the $z^\prime$-band
data as the fluxes at $i^\prime$-band are affected by Ly$\alpha$ emission. 
For objects fainter than $1 \sigma$ in the $z^\prime$ band, 
we calculate the upper-limit of the UV luminosity, $L_\nu ({\rm UV})$, and 
the lower-limit of the rest-frame equivalent width, $EW_0({\rm Ly \alpha})$. 
The $EW_0({\rm Ly}\alpha)$ distribution is similar to those in previous 
studies of LAEs at $z \sim 3$--6 
(e.g., Shimasaku et al. 2006; Dawson et al. 2007; Ouchi et al. 2008; Gronwall et al. 2008), 
with the mean rest-frame Lyman $\alpha$ equivalent widths of the sample smaller than 200 \AA. 
There is no LAE with $EW_0({\rm Ly}\alpha) > 250$ \AA~in our sample, 
although the rest-frame Ly$\alpha$ equivalent widths of 23 of the 
LAEs in our sample (29\%) are lower limits. 
Taking account of a predicted $EW_0({\rm Ly}\alpha)$ for starburst galaxies, 
300 \AA~for young starburst (age $\le 10^6$ yr) and 
100 \AA~for old starburst (age $\sim 10^8$ yr) (Malhotra \& Rhoads 2002), 
we consider that there is no peculiar object in our sample. 
Figure \ref{La:ewMuv} shows the relation between $EW_0({\rm Ly}\alpha)$ and $M_{\rm UV}$. 
There is no object with $EW_0({\rm Ly}\alpha) > 80$\AA~ in the 
UV-bright ($M_{\rm UV} < -21.5$) sample. 
Although the number of UV-bright LAEs is small and the uncertainties on $EW_0$s for 
UV-faint objects are large, this trend is similar to that found for LBGs and LAEs at 
$z \sim 5$--6 (Ando et al. 2006; Shimasaku et al. 2006; Ouchi et al. 2008). 
We conclude that our sample shows the ``average'' picture of bright LAEs at $z \sim 5$. 

\subsection{UV Luminosity Function}

Figure \ref{La:UVLF} shows UV LF of our sample (black symbols). 
The UV LFs of LAEs estimated in the four subfields are consistent within errors. 
Figure \ref{La:UVLF} also include the UV LF of LBGs at $z \sim 5$ (Yoshida et al. 2006) 
and LAEs at $z \sim 3.1$, 3.7, and 5.7 (Ouchi et al. 2008). 
The shape of our UV LF seems different from those of previous works and is not fit by 
Schechter function, since 
a detection limit of rest-frame equivalent width, $EW_0({\rm Ly}\alpha)$, 
depends on $M_{UV}$, e.g., $EW_0({\rm Ly}\alpha) > 11$\AA~ at $M_{\rm UV}<-21.5$ and 
$EW_0({\rm Ly}\alpha)>57$\AA~ at $M_{\rm UV}=-20$ (see Figure \ref{La:ewMuv}). 
As a reference, we overlay the result of our Monte Carlo simulation for $\alpha=-1.5$ and 
$\sigma_{EW}=100$ \AA: dotted line show input UV LF for $EW_0({\rm Ly}\alpha) > 11$\AA~ 
and solid line show output UV LF. This result also shows that our UV LF is considered 
to be complete for LAE with $EW_0 > 11$\AA~ for $M_{\rm UV} < -21.5$. 
We therefore concentrate the number density at $M_{\rm UV} < -21.5$. 

First, we compare our UV LF with that of LBGs at $z \sim 5$. 
The number density of our LAEs is comparable to that of LBGs at $z \sim 5$ at 
$M_{\rm UV} \sim -22$ and $\sim 20$--25 \% at $M_{\rm UV} = -21.5$. 
Ouchi et al. (2008) pointed out that the ratio of number densities of LAEs to 
those of LBGs is $\sim 10$\% at $z = 3$--4 and $>50$\% at $z = 5.7$. 
Our result imply that the ratio of the number density of LAEs to that of LBGs 
becomes larger with redshift from $z = 4$ to 5. 
Next, we compare our UV LF with those of LAEs at different redshifts. 
Figure \ref{La:UVLF2} shows the number density of LAEs at $M_{\rm UV}=-21.5$ 
as a function of $z$. 
The number density of our LAEs at $M_{\rm UV} = -21.5$ is comparable to that of 
LAEs at $z \sim 5.7$, while larger than those of LAEs at $z \sim 3.1$ and 3.7. 
The number density of UV-bright LAEs ($M_{\rm UV}<-21.5$) increases an order of 
magnitude with redshift from $z=4$ to $5$. 
Since it is likely that the LAEs are star-forming galaxies in an earlier star formation phase, 
our findings imply that the initial active star-formation phase occur mainly 
beyond $z=5$. 

\subsection{Clustering Properties}

We found the large scale structure of LAEs of $0.4^\circ \times 0.2^\circ$ 
in subsection \ref{sec:SD}. 
In order to perform a more quantitative study of the 
clustering properties of the LAEs at $z\sim 4.86$, 
we derive their angular two-point correlation function (ACF), $w(\theta)$, 
using the estimator defined by Landy \& Szalay (1993), 
\begin{equation}
 w(\theta) = \frac{DD(\theta)-2DR(\theta)+RR(\theta)}{RR(\theta)},
 \label{two-point}
\end{equation}
where $DD(\theta)$, $DR(\theta)$, and $RR(\theta)$ are normalized numbers 
of galaxy-galaxy, galaxy-random, and random-random pairs, respectively. 
The random sample here consists of 100,000 sources with the same geometrical 
constraints as the galaxy sample. The observed ACF is fit well by a single power law: 
$w(\theta)=0.021^{+0.025}_{-0.011}\,\theta^{-0.90\pm 0.22}$ (Figure \ref{La:acf}). 
The correlation length, $r_0$, is calculated 
from the ACF through Limber's equation (e.g., Peebles 1980), 
assuming a top-hat redshift distribution centered on $z=4.86\pm 0.03$. 
We estimated the $r_0$ corresponding to our sample of LAEs 
as $r_0 = 4.4^{+5.7}_{-2.9}$ Mpc.  
The two-point correlation function is thus written as 
$\xi(r) = (r/4.4^{+5.7}_{-2.9}\,{\rm Mpc})^{-1.90\pm 0.22}$. 
This agrees well with results from other works at similar redshifts, e.g., 
$r_0=5.0\pm 0.4$ for $z\simeq 4.9$ (Ouchi et al. 2003); 
$r_0=4.57\pm 0.60$ for $z\simeq 4.5$ (Kova\v c et al. 2007). 

Also shown in Figure \ref{La:acf} are the ACFs for the LAEs in the four subfields. 
We detect strong clustering signals in small scale ($\theta\le 50$ arcsec) for NE, 
SW, and SE subfields, with the NW subfield showing no clustering signals at any 
angular separations. 
Although this may imply the presence of a cosmic variance on the clustering properties similar to that found 
in a previous study (Shimasaku et al. 2004), 
taking account of large uncertainties of ACFs, 
we consider that there are no significant field-to-field variations 
among the four subfields. 

\section{SUMMARY}

We have performed the largest survey to date for Ly$\alpha$ emitters at $z \approx 4.86$, 
using narrow-band (NB711) imaging technique in the COSMOS 2 square degree field. 
We have found a total of 79 Ly$\alpha$ emission-line galaxy candidates. 
For 7 LAE candidates with available spectroscopic data, we have confirmed 
that our criteria for selecting LAEs at $z \approx 4.86$ are working well. 
Our results and conclusions are summarized below, 

1. We have found a field-to-field variation of the number density of LAEs 
as large as a factor of $\simeq 2$ among the nine subfields with $0.5^\circ \times 0.5^\circ$. 
On the other hand, the number density of LAEs for four subfields 
with $0.7^\circ \times 0.7^\circ$ is consistent within a error. 
This finding is consistent with the scale of large scale structure we found, 
50 Mpc $\times$ 25 Mpc. 
We conclude that at least 0.5 deg$^2$ survey area is required 
to derive averaged properties of LAEs at $z\sim 5$, and our survey field is wide enough 
to overcome the cosmic variance. 

2. The Ly$\alpha$ LF is well-fitted by 
a Schechter function with best-fit Schechter parameters: 
$\log L^*=42.91^{+0.49}_{-0.31}{\rm ergs \; s^{-1}}$ 
and $\phi^* = 1.22^{+8.02}_{-1.05} \times 10^{-4}{\rm Mpc^{-3}}$ for $\alpha=-1.5$ (fixed). 
The two-point correlation function is well fitted 
by a power law, $w(\theta) = 0.021^{+0.025}_{-0.011}\theta^{-0.90\pm0.22}$, 
giving $\xi(r)=(r/4.4^{+5.7}_{-2.9} {\rm Mpc})^{-1.9}$. 

3. We have derived the UV LF of LAEs. 
The number density of our LAEs at $M_{UV}=-21.5$ are similar to those of LAEs at 
$z \sim 5.7$ while larger than those of LAEs at $z \sim 3$--4. 
The number density of UV-bright LAEs increases an order of magnitude with redshift from $z \sim 4$ to $z \sim 5$. 

\acknowledgements
The HST COSMOS Treasury program was supported through NASA grant HST-GO-09822.
We greatly acknowledge the contributions of the entire COSMOS collaboration
consisting of more than 70 scientists. 
We would like to thank Masami Ouchi for useful discussion 
and the anonymous referee for valuable comments. 
We would also like to thank the Subaru Telescope staff for their invaluable help. 
This work was financially supported in part by the JSPS (Nos. 15340059 and 17253001).


\clearpage
\pagestyle{empty}
%
%
\begin{deluxetable}{llcccccccccc}
\tablecaption{\label{La:tab:cover}A list of Ly$\alpha$ emitter candidates.}
\tablewidth{0pt}
\tablehead{
\colhead{\#} &
\colhead{RA} &
\colhead{DEC} &
\colhead{$r^\prime$} &
\colhead{$i^\prime$} &
\colhead{$ri$} & 
\colhead{$NB711$} &
\colhead{$z^\prime$} & 
\colhead{$\log L({\rm Ly}\alpha)$} &
\colhead{$\log L_\nu$(1540\AA)} &
\colhead{$M_{UV}$} &
\colhead{$EW_0$}
\\
\colhead{} &
\colhead{(deg)} &
\colhead{(deg)} &
\colhead{(mag)} &
\colhead{(mag)} &
\colhead{(mag)} &
\colhead{(mag)} &
\colhead{(mag)} &
\colhead{($\rm ergs \; s^{-1}$)} & 
\colhead{($\rm ergs \; s^{-1} \; Hz^{-1}$)} & 
\colhead{(mag)} &
\colhead{(\AA)}
}
\startdata
  1 & 150.68983 & 1.598039 & 28.85 & 26.54 & 26.87 & 24.86 &$  25.95$ & 42.68 &$  28.80$ &$ -20.4$&$   46$\\
  2 & 150.58466 & 1.528353 & 26.26 & 25.13 & 25.36 & 23.52 &$  25.40$ & 43.20 &$  29.02$ &$ -21.0$&$   90$\\
  3 & 150.43377 & 1.584748 & 27.49 & 26.38 & 26.61 & 24.46 &$> 26.64$ & 42.85 &$< 28.52$ &$>-19.7$&$> 127$\\
  4 & 150.47017 & 1.527121 & 27.85 & 26.69 & 26.92 & 24.76 &$  26.28$ & 42.73 &$  28.67$ &$ -20.1$&$   69$\\
  5 & 150.12679 & 1.606008 & 27.15 & 25.63 & 25.91 & 23.43 &$  25.61$ & 43.28 &$  28.94$ &$ -20.7$&$  131$\\
  6 & 150.19137 & 1.514911 & 25.95 & 25.13 & 25.32 & 24.47 &$  24.79$ & 42.64 &$  29.26$ &$ -21.6$&$   14$\\
  7 & 149.42627 & 1.570369 & 27.34 & 25.94 & 26.21 & 24.44 &$> 26.64$ & 42.83 &$< 28.52$ &$>-19.7$&$> 121$\\
  8 & 149.50750 & 1.569846 & 26.40 & 24.72 & 25.01 & 24.05 &$  25.06$ & 42.85 &$  29.15$ &$ -21.3$&$   30$\\
  9 & 150.72870 & 1.654431 & 27.26 & 26.02 & 26.27 & 24.22 &$> 26.64$ & 42.94 &$< 28.52$ &$>-19.7$&$> 156$\\
 10 & 150.69867 & 1.658967 & 99.00 & 27.04 & 27.42 & 24.80 &$> 26.64$ & 42.74 &$< 28.52$ &$>-19.7$&$>  98$\\
 11 & 150.69845 & 1.643227 & 27.19 & 25.73 & 26.00 & 24.55 &$  24.98$ & 42.75 &$  29.19$ &$ -21.4$&$   22$\\
 12 & 150.44771 & 1.639259 & 26.33 & 24.16 & 24.48 & 23.64 &$  24.31$ & 42.98 &$  29.46$ &$ -22.0$&$   20$\\
 13 & 150.21979 & 1.647579 & 27.64 & 25.97 & 26.26 & 24.71 &$  25.67$ & 42.70 &$  28.91$ &$ -20.7$&$   36$\\
 14 & 149.92387 & 1.706955 & 27.91 & 26.99 & 27.19 & 24.22 &$> 26.64$ & 42.98 &$< 28.52$ &$>-19.7$&$> 171$\\
 15 & 149.86911 & 1.741172 & 27.83 & 25.18 & 25.52 & 24.07 &$  25.43$ & 42.94 &$  29.01$ &$ -20.9$&$   51$\\
 16 & 149.81740 & 1.738043 & 26.33 & 25.43 & 25.63 & 24.46 &$  25.87$ & 42.73 &$  28.83$ &$ -20.5$&$   47$\\
 17 & 149.85339 & 1.702846 & 28.17 & 26.76 & 27.03 & 24.74 &$> 26.64$ & 42.75 &$< 28.52$ &$>-19.7$&$> 100$\\
 18 & 149.81761 & 1.638761 & 26.26 & 25.23 & 25.45 & 24.10 &$  25.25$ & 42.91 &$  29.08$ &$ -21.1$&$   41$\\
 19 & 149.47913 & 1.713145 & 26.26 & 24.90 & 25.17 & 24.18 &$  24.82$ & 42.81 &$  29.25$ &$ -21.5$&$   21$\\
 20 & 150.77783 & 1.795379 & 27.63 & 25.85 & 26.15 & 24.38 &$  25.44$ & 42.85 &$  29.00$ &$ -20.9$&$   42$\\
 21 & 150.43713 & 1.821238 & 27.16 & 25.26 & 25.57 & 23.96 &$  24.83$ & 43.00 &$  29.25$ &$ -21.5$&$   34$\\
 22 & 150.39265 & 1.852772 & 30.47 & 26.36 & 26.74 & 24.51 &$> 26.64$ & 42.84 &$< 28.52$ &$>-19.7$&$> 122$\\
 23 & 150.31602 & 1.848847 & 26.35 & 24.94 & 25.21 & 24.41 &$  24.73$ & 42.65 &$  29.29$ &$ -21.6$&$   14$\\
 24 & 149.98396 & 1.914333 & 27.61 & 26.12 & 26.39 & 24.49 &$> 26.64$ & 42.82 &$< 28.52$ &$>-19.7$&$> 119$\\
 25 & 149.76505 & 1.835950 & 26.46 & 25.04 & 25.31 & 24.38 &$  24.49$ & 42.71 &$  29.38$ &$ -21.9$&$   13$\\
 26 & 149.82192 & 1.826156 & 27.56 & 26.12 & 26.39 & 24.46 &$> 26.64$ & 42.83 &$< 28.52$ &$>-19.7$&$> 122$\\
 27 & 149.70144 & 1.880336 & 29.66 & 25.88 & 26.26 & 24.38 &$  25.73$ & 42.86 &$  28.89$ &$ -20.6$&$   56$\\
 28 & 149.43575 & 1.958916 & 25.53 & 24.38 & 24.62 & 23.34 &$  24.16$ & 43.20 &$  29.51$ &$ -22.2$&$   29$\\
 29 & 150.52280 & 2.053999 & 28.66 & 26.82 & 27.13 & 23.87 &$> 26.64$ & 43.13 &$< 28.52$ &$>-19.7$&$> 240$\\
 30 & 150.44155 & 2.045647 & 25.92 & 24.17 & 24.47 & 23.50 &$  24.13$ & 43.07 &$  29.53$ &$ -22.2$&$   21$\\
 31 & 149.80258 & 1.976421 & 28.37 & 26.13 & 26.46 & 24.58 &$  25.48$ & 42.78 &$  28.99$ &$ -20.9$&$   37$\\
 32 & 149.47068 & 2.112708 & 28.53 & 27.11 & 27.38 & 24.84 &$> 26.64$ & 42.72 &$< 28.52$ &$>-19.7$&$>  94$\\
 33 & 149.49438 & 2.111401 & 27.52 & 26.13 & 26.39 & 24.64 &$> 26.64$ & 42.75 &$< 28.52$ &$>-19.7$&$> 100$\\
 34 & 149.50653 & 2.059920 & 26.21 & 24.78 & 25.05 & 23.56 &$  25.14$ & 43.15 &$  29.12$ &$ -21.2$&$   63$\\
 35 & 149.42625 & 1.971732 & 27.00 & 25.60 & 25.87 & 23.59 &$  25.36$ & 43.21 &$  29.04$ &$ -21.0$&$   89$\\
 36 & 150.75362 & 2.237688 & 27.53 & 26.18 & 26.45 & 24.37 &$  26.27$ & 42.88 &$  28.67$ &$ -20.1$&$   97$\\
 37 & 150.74435 & 2.216502 & 25.70 & 24.59 & 24.82 & 24.08 &$  24.31$ & 42.76 &$  29.45$ &$ -22.0$&$   12$\\
 38 & 150.23097 & 2.219221 & 27.93 & 26.37 & 26.65 & 24.04 &$> 26.64$ & 43.04 &$< 28.52$ &$>-19.7$&$> 196$\\
 39 & 150.17687 & 2.162903 & 27.02 & 26.05 & 26.26 & 24.21 &$> 26.64$ & 42.95 &$< 28.52$ &$>-19.7$&$> 157$\\
 40 & 149.96795 & 2.258172 & 27.64 & 26.05 & 26.34 & 24.76 &$  26.26$ & 42.68 &$  28.68$ &$ -20.1$&$   60$\\
 41 & 150.01739 & 2.146056 & 26.51 & 24.88 & 25.17 & 23.39 &$  25.62$ & 43.25 &$  28.93$ &$ -20.7$&$  124$\\
 42 & 149.83435 & 2.270296 & 26.92 & 25.37 & 25.65 & 23.82 &$  25.58$ & 43.08 &$  28.95$ &$ -20.8$&$   81$\\
 43 & 150.68548 & 2.422582 & 26.74 & 25.83 & 26.03 & 24.49 &$  25.59$ & 42.78 &$  28.94$ &$ -20.8$&$   41$\\
 44 & 150.48986 & 2.405317 & 26.14 & 24.93 & 25.17 & 23.91 &$  25.16$ & 42.97 &$  29.11$ &$ -21.2$&$   43$\\
 45 & 150.34351 & 2.380535 & 26.92 & 26.01 & 26.22 & 24.62 &$  26.54$ & 42.74 &$  28.56$ &$ -19.8$&$   89$\\
 46 & 150.17116 & 2.443712 & 27.53 & 25.48 & 25.79 & 24.65 &$  25.27$ & 42.66 &$  29.07$ &$ -21.1$&$   23$\\
 47 & 149.95843 & 2.414291 & 29.20 & 26.60 & 26.95 & 24.70 &$  26.49$ & 42.76 &$  28.58$ &$ -19.9$&$   89$\\
 48 & 149.86004 & 2.390346 & 27.06 & 26.09 & 26.31 & 23.95 &$> 26.64$ & 43.07 &$< 28.52$ &$>-19.7$&$> 208$\\
 49 & 149.62681 & 2.428601 & 31.96 & 26.72 & 27.11 & 24.93 &$  25.84$ & 42.66 &$  28.84$ &$ -20.5$&$   40$\\
 50 & 149.51027 & 2.301385 & 27.33 & 26.22 & 26.45 & 23.69 &$  26.34$ & 43.19 &$  28.64$ &$ -20.0$&$  208$\\
 51 & 150.72903 & 2.584166 & 26.45 & 25.56 & 25.76 & 24.64 &$  25.67$ & 42.65 &$  28.91$ &$ -20.7$&$   33$\\
 52 & 150.78495 & 2.573355 & 26.58 & 25.42 & 25.66 & 24.62 &$  25.31$ & 42.64 &$  29.06$ &$ -21.0$&$   23$\\
 53 & 150.75115 & 2.481606 & 28.43 & 26.25 & 26.57 & 24.41 &$> 26.64$ & 42.87 &$< 28.52$ &$>-19.7$&$> 133$\\
 54 & 150.24314 & 2.530345 & 27.07 & 25.42 & 25.71 & 23.45 &$  25.44$ & 43.26 &$  29.00$ &$ -20.9$&$  108$\\
 55 & 150.13505 & 2.486044 & 26.83 & 25.21 & 25.50 & 24.50 &$  25.22$ & 42.68 &$  29.09$ &$ -21.1$&$   23$\\
 56 & 149.89657 & 2.527743 & 26.57 & 25.42 & 25.66 & 24.45 &$  26.08$ & 42.75 &$  28.75$ &$ -20.3$&$   59$\\
 57 & 149.75765 & 2.572967 & 28.44 & 26.53 & 26.84 & 24.67 &$  25.54$ & 42.77 &$  28.96$ &$ -20.8$&$   38$\\
 58 & 149.87223 & 2.497300 & 27.31 & 25.72 & 26.00 & 23.90 &$  25.50$ & 43.07 &$  28.98$ &$ -20.9$&$   74$\\
 59 & 149.60335 & 2.612591 & 26.85 & 25.52 & 25.78 & 24.52 &$  26.03$ & 42.73 &$  28.77$ &$ -20.3$&$   55$\\
 60 & 149.58816 & 2.521003 & 27.14 & 25.43 & 25.73 & 24.14 &$  25.28$ & 42.93 &$  29.07$ &$ -21.1$&$   43$\\
 61 & 149.46094 & 2.563734 & 27.24 & 26.09 & 26.33 & 24.80 &$  25.71$ & 42.66 &$  28.90$ &$ -20.6$&$   35$\\
 62 & 150.80329 & 2.730062 & 26.73 & 25.50 & 25.75 & 24.59 &$  25.90$ & 42.68 &$  28.82$ &$ -20.5$&$   43$\\
 63 & 150.76442 & 2.688660 & 25.39 & 24.31 & 24.53 & 23.68 &$  23.91$ & 42.96 &$  29.62$ &$ -22.4$&$   13$\\
 64 & 150.43973 & 2.720629 & 26.24 & 25.09 & 25.33 & 24.27 &$  25.30$ & 42.78 &$  29.06$ &$ -21.1$&$   32$\\
 65 & 150.30282 & 2.772591 & 26.65 & 25.06 & 25.34 & 24.41 &$  24.81$ & 42.70 &$  29.26$ &$ -21.5$&$   16$\\
 66 & 150.30023 & 2.666173 & 28.11 & 26.84 & 27.09 & 24.88 &$> 26.64$ & 42.69 &$< 28.52$ &$>-19.7$&$>  87$\\
 67 & 150.28152 & 2.651694 & 27.53 & 26.16 & 26.42 & 24.16 &$> 26.64$ & 42.98 &$< 28.52$ &$>-19.7$&$> 170$\\
 68 & 150.29721 & 2.634812 & 26.25 & 24.65 & 24.94 & 24.06 &$  25.12$ & 42.82 &$  29.13$ &$ -21.2$&$   29$\\
 69 & 149.94445 & 2.704370 & 26.02 & 25.20 & 25.39 & 24.49 &$  25.79$ & 42.66 &$  28.86$ &$ -20.6$&$   37$\\
 70 & 149.78731 & 2.678302 & 26.12 & 24.67 & 24.94 & 23.50 &$  24.88$ & 43.16 &$  29.23$ &$ -21.5$&$   52$\\
 71 & 149.66107 & 2.739789 & 26.23 & 24.87 & 25.14 & 24.32 &$  24.66$ & 42.69 &$  29.31$ &$ -21.7$&$   14$\\
 72 & 149.72632 & 2.664706 & 99.00 & 25.59 & 25.98 & 24.35 &$> 26.64$ & 42.85 &$< 28.52$ &$>-19.7$&$> 126$\\
 73 & 149.43064 & 2.784033 & 28.95 & 26.30 & 26.65 & 24.81 &$> 26.64$ & 42.69 &$< 28.52$ &$>-19.7$&$>  87$\\
 74 & 149.41782 & 2.735198 & 26.83 & 25.45 & 25.71 & 24.41 &$> 26.64$ & 42.78 &$< 28.52$ &$>-19.7$&$> 107$\\
 75 & 149.44796 & 2.694757 & 26.55 & 25.36 & 25.60 & 24.54 &$  24.89$ & 42.68 &$  29.22$ &$ -21.5$&$   17$\\
 76 & 150.31928 & 2.864155 & 27.78 & 26.37 & 26.64 & 24.66 &$> 26.64$ & 42.76 &$< 28.52$ &$>-19.7$&$> 102$\\
 77 & 150.23253 & 2.849228 & 99.00 & 26.19 & 26.58 & 24.70 &$> 26.64$ & 42.74 &$< 28.52$ &$>-19.7$&$>  97$\\
 78 & 149.80660 & 2.861745 & 25.88 & 25.05 & 25.24 & 23.25 &$  24.83$ & 43.33 &$  29.25$ &$ -21.5$&$   71$\\
 79 & 149.43851 & 2.902153 & 26.54 & 25.58 & 25.79 & 24.62 &$  25.48$ & 42.67 &$  28.99$ &$ -20.9$&$   29$\\
\enddata                                                                                  

\end{deluxetable}

\clearpage

\begin{deluxetable}{lccc}
\tablecaption{\label{La:tab:CV}The number density of LAEs in the 9 subfields of $0.5^\circ \times 0.5^\circ$.}
\tablewidth{0pt}
\tablehead{
\colhead{} & 
\colhead{East} & 
\colhead{Middle} &
\colhead{West} \\
\colhead{} & 
\colhead{(deg$^{-2}$)} & 
\colhead{(deg$^{-2}$)} &
\colhead{(deg$^{-2}$)}
}
\startdata
North  & $29.6 \pm 12.1$ & $52.0 \pm 15.7$ & $64.3 \pm 17.8$ \\
Middle & $30.9 \pm 12.6$ & $28.0 \pm 11.4$ & $45.3 \pm 15.1$ \\
South  & $59.5 \pm 17.9$ & $28.0 \pm 11.4$ & $53.1 \pm 16.0$ \\
\enddata                                                                                  

\end{deluxetable}


\begin{deluxetable}{ccc}
\tablecaption{\label{La:tab:LALF}Best-fit Schechter parameters for Ly$\alpha$ luminosity functions.}
\tablewidth{0pt}
\tablehead{
\colhead{$\alpha$} & 
\colhead{$\log L^*_{\rm Ly \alpha}$} &
\colhead{$\phi^{*}$} 
\\
\colhead{(fixed)} &
\colhead{(erg s$^{-1}$)} &
\colhead{($\times 10^{-4} {\rm Mpc}^{-3}$)} 
}
\startdata
$-1.0$ & $42.82^{+0.39}_{-0.28}$ & $1.41^{+6.73}_{-1.09}$ \\
$-1.5$ & $42.91^{+0.49}_{-0.31}$ & $1.22^{+8.02}_{-1.05}$ \\
$-2.0$ & $43.00^{+0.70}_{-0.37}$ & $0.82^{+9.98}_{-0.77}$ \\
\enddata                                                                                  

\end{deluxetable}


\begin{deluxetable}{llcc}
\tablecaption{\label{La:tab:LALF2}Best-fit Schechter parameters for Ly$\alpha$ luminosity functions for each subfield.}
\tablewidth{0pt}
\tablehead{
\colhead{$\alpha$} & 
\colhead{subfield} & 
\colhead{$\log L^*_{\rm Ly \alpha}$} &
\colhead{$\phi^{*}$} 
\\
\colhead{(fixed)} &
\colhead{} &
\colhead{(erg s$^{-1}$)} &
\colhead{($\times 10^{-4} {\rm Mpc}^{-3}$)} 
}
\startdata
-1.0&NE & 42.79& 0.85\\
    &NW & 42.76& 3.0 \\
    &SW & 42.92& 0.84\\
    &SE & 42.83& 1.4 \\
\hline
-1.5&NE & 42.87& 0.78\\
    &NW & 42.84& 2.7 \\
    &SW & 43.02& 0.68\\
    &SE & 42.91& 1.2 \\
\hline
-2.0&NE & 42.97& 0.56\\
    &NW & 42.94& 2.0 \\
    &SW & 43.15& 0.40\\
    &SE & 43.02& 0.83\\
\enddata
\end{deluxetable}

\clearpage


\begin{figure}
\epsscale{0.8}
\plotone{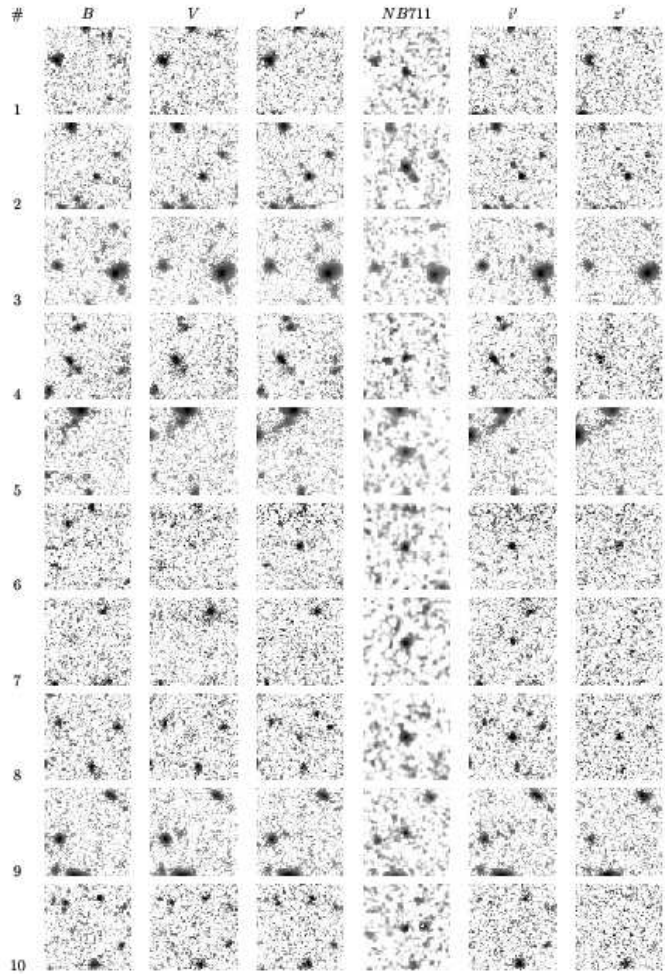}
\caption{
Broad-band and NB711 images of our LAE candidates at $z \approx 4.9$.
Each box is $12^{\prime \prime}$ on a side (north is up and east is left).
\label{thum}
}
\end{figure}
\clearpage
{\plotone{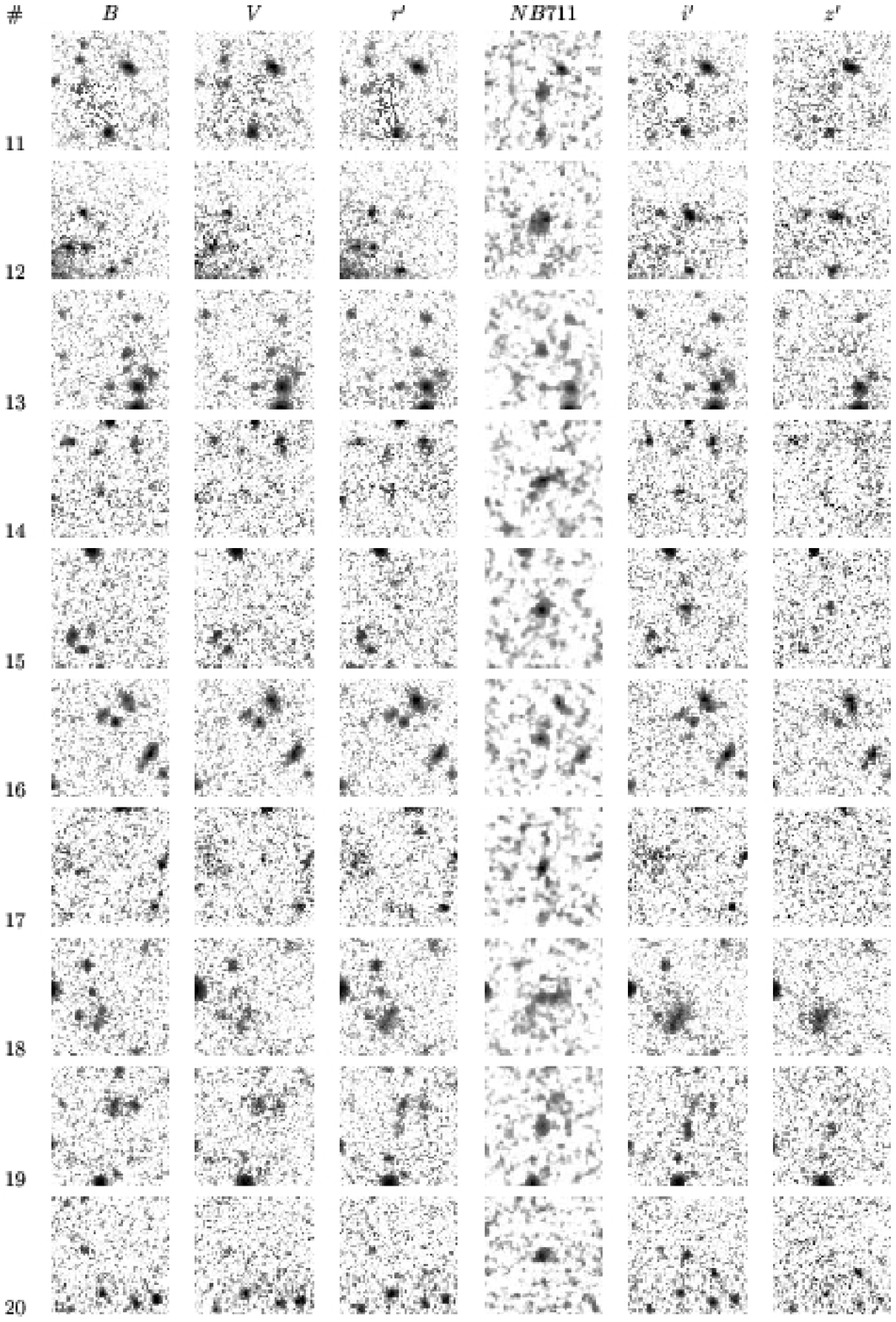}}\\[2mm]
\centerline{Fig. 4b. --- continued.}
\clearpage
{\plotone{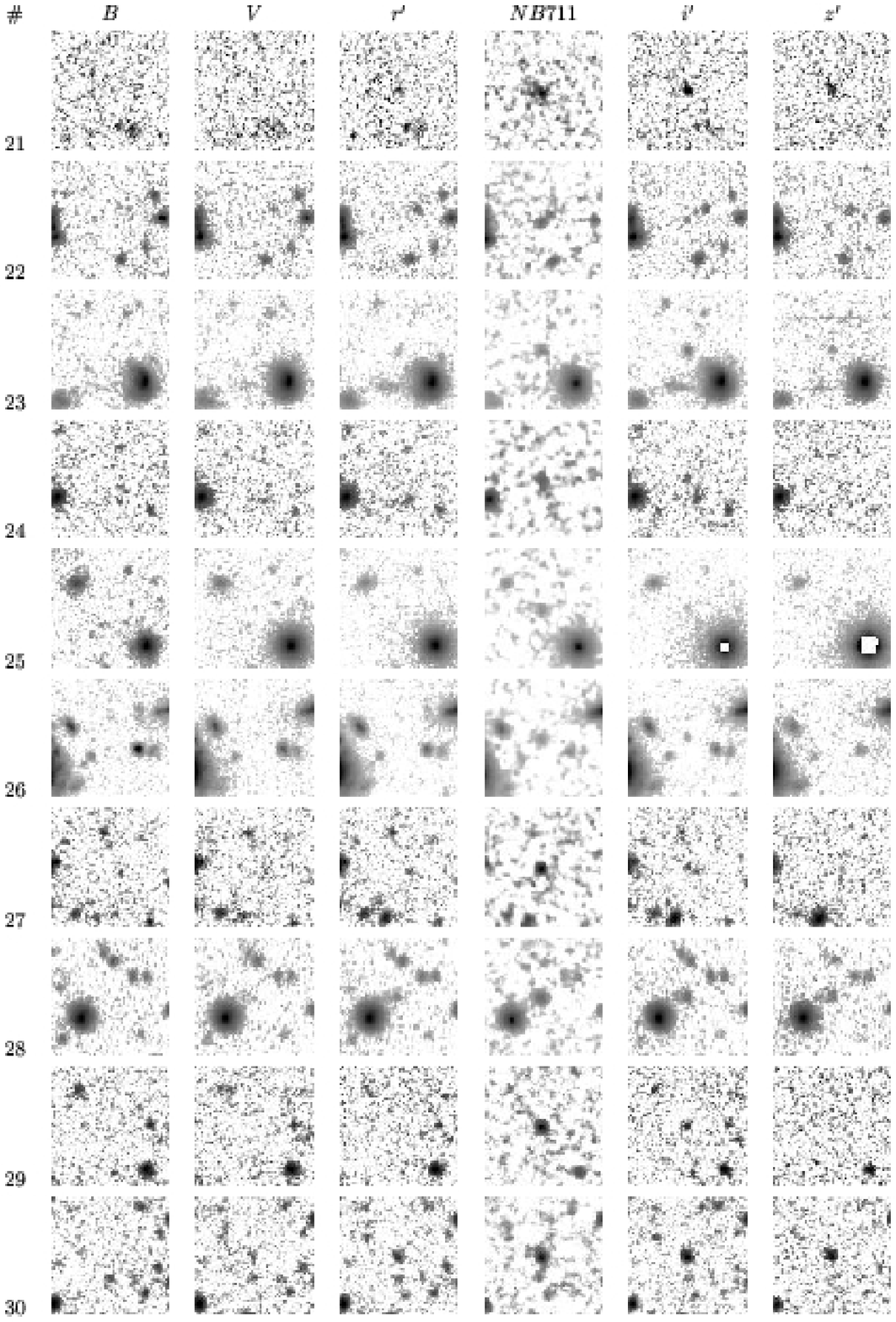}}\\[2mm]
\centerline{Fig. 4c. --- continued.}
\clearpage
{\plotone{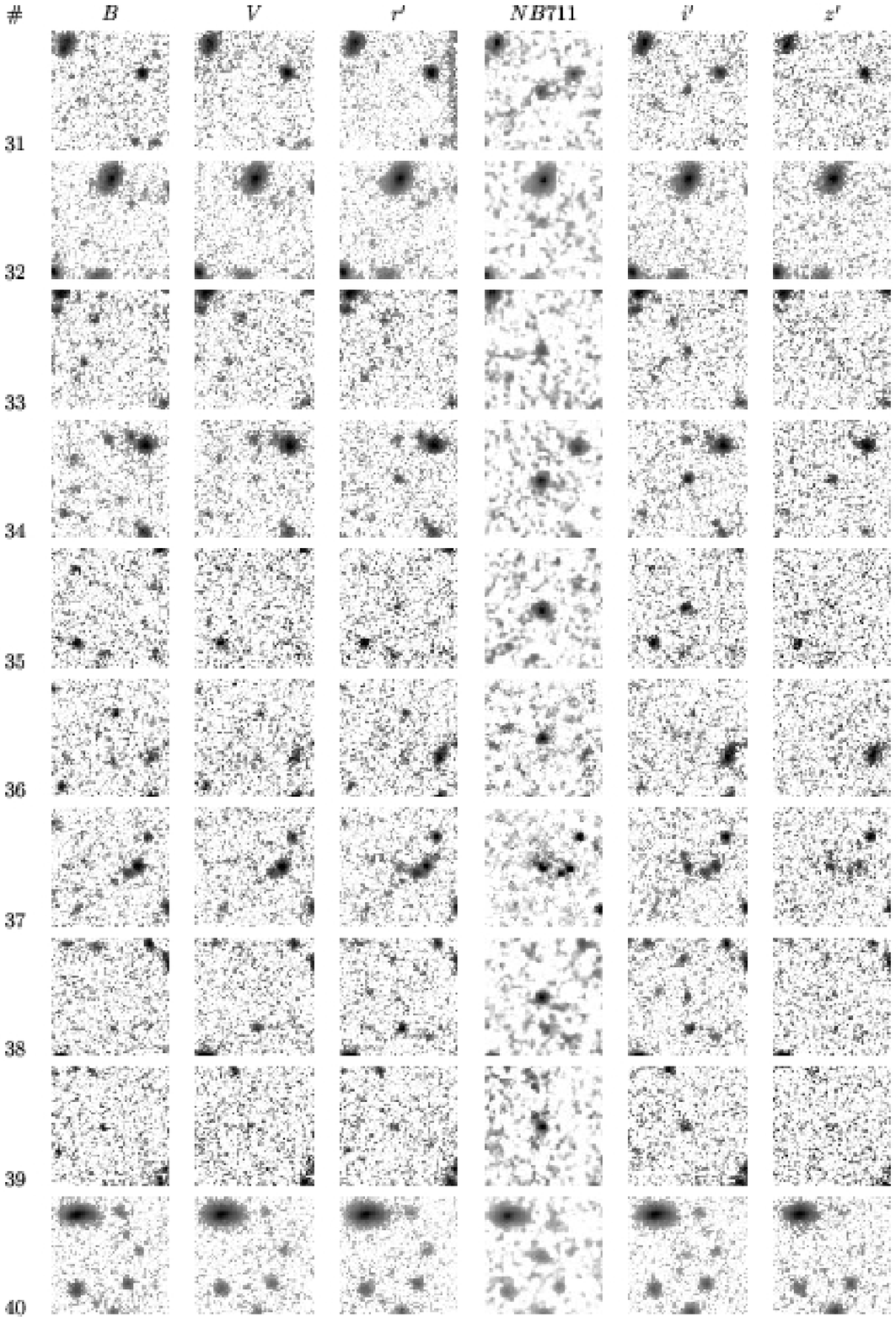}}\\[2mm]
\centerline{Fig. 4d. --- continued.}
\clearpage
{\plotone{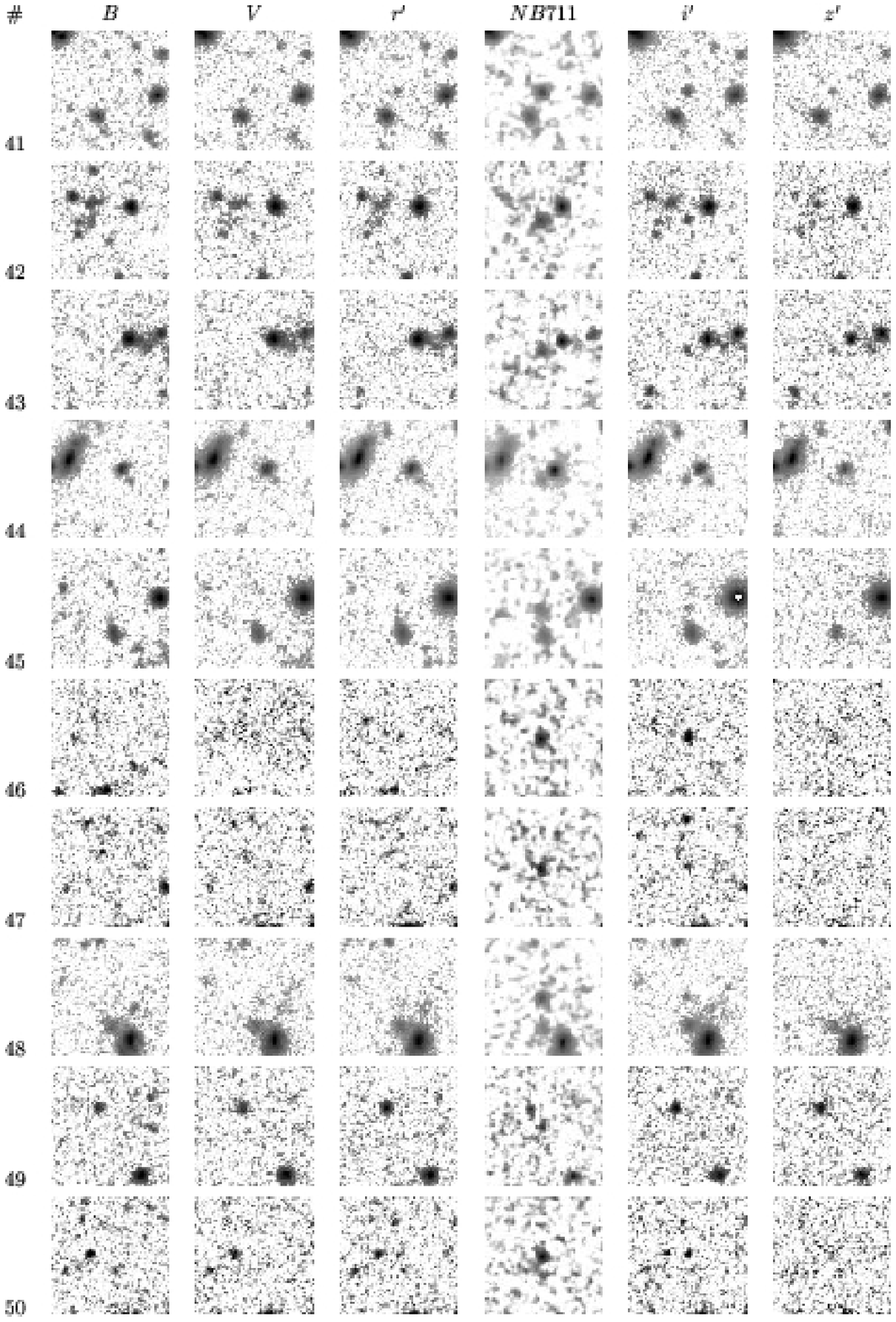}}\\[2mm]
\centerline{Fig. 4e. --- continued.}
\clearpage
{\plotone{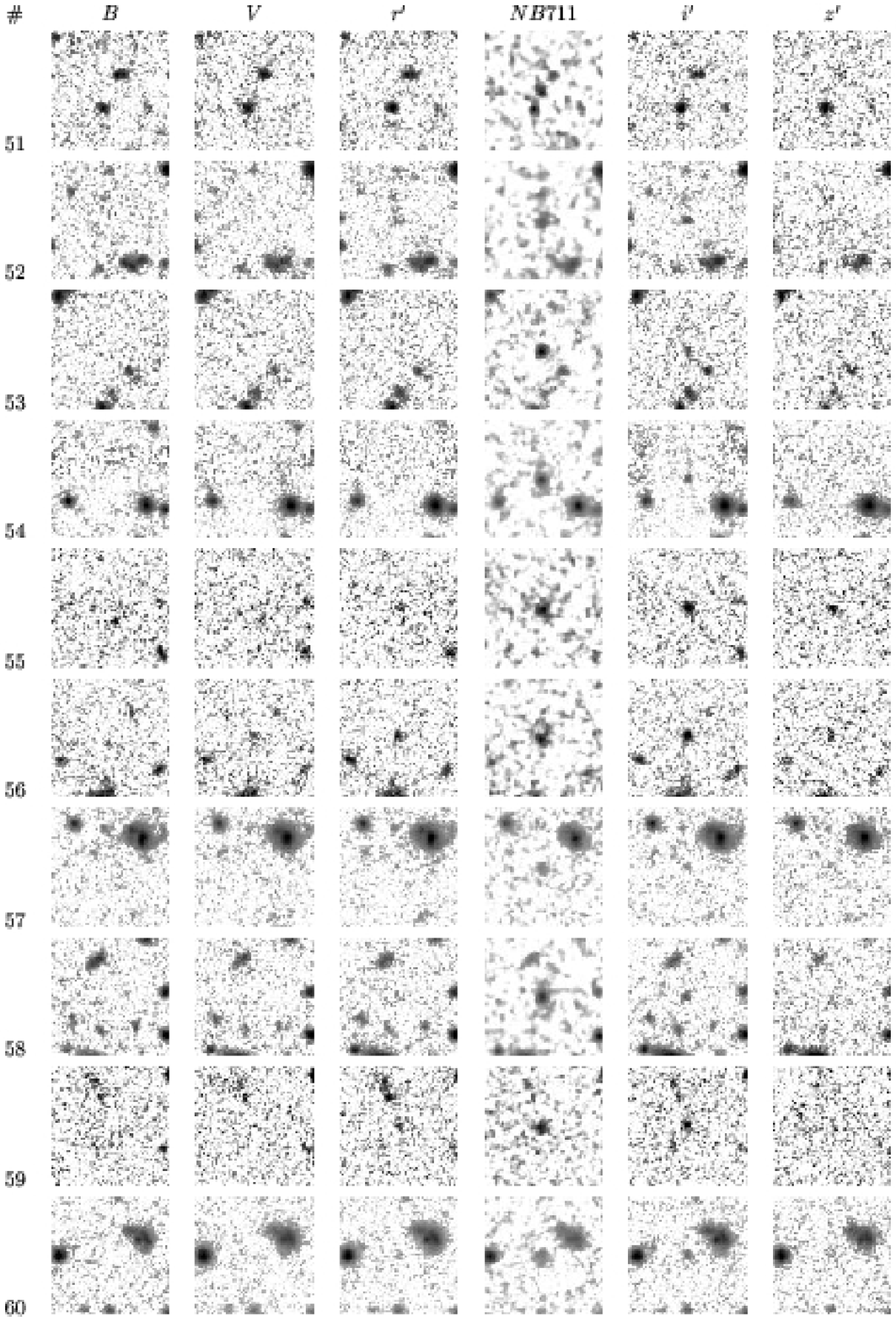}}\\[2mm]
\centerline{Fig. 4f. --- continued.}
\clearpage
{\plotone{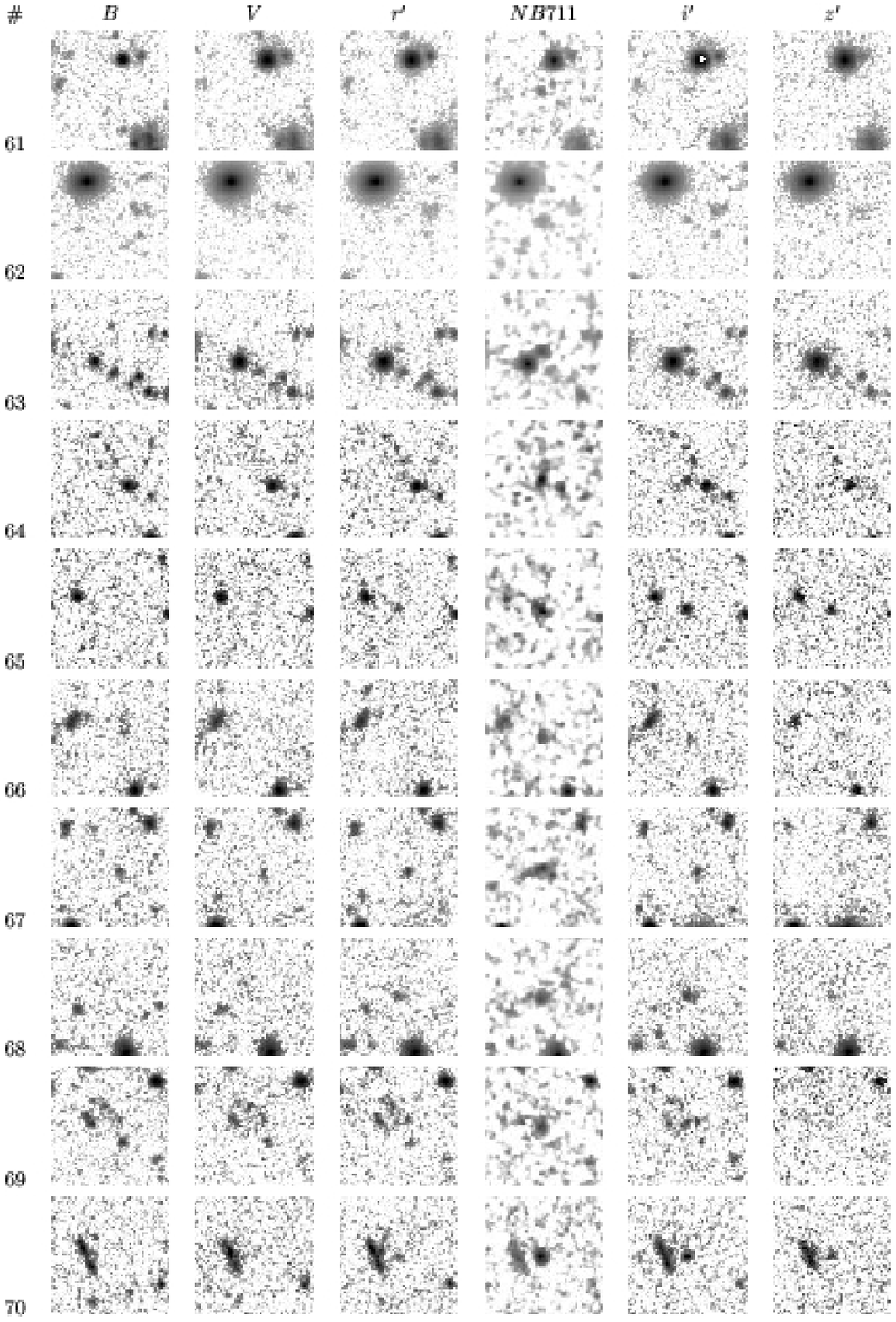}}\\[2mm]
\centerline{Fig. 4g. --- continued.}
\clearpage
{\plotone{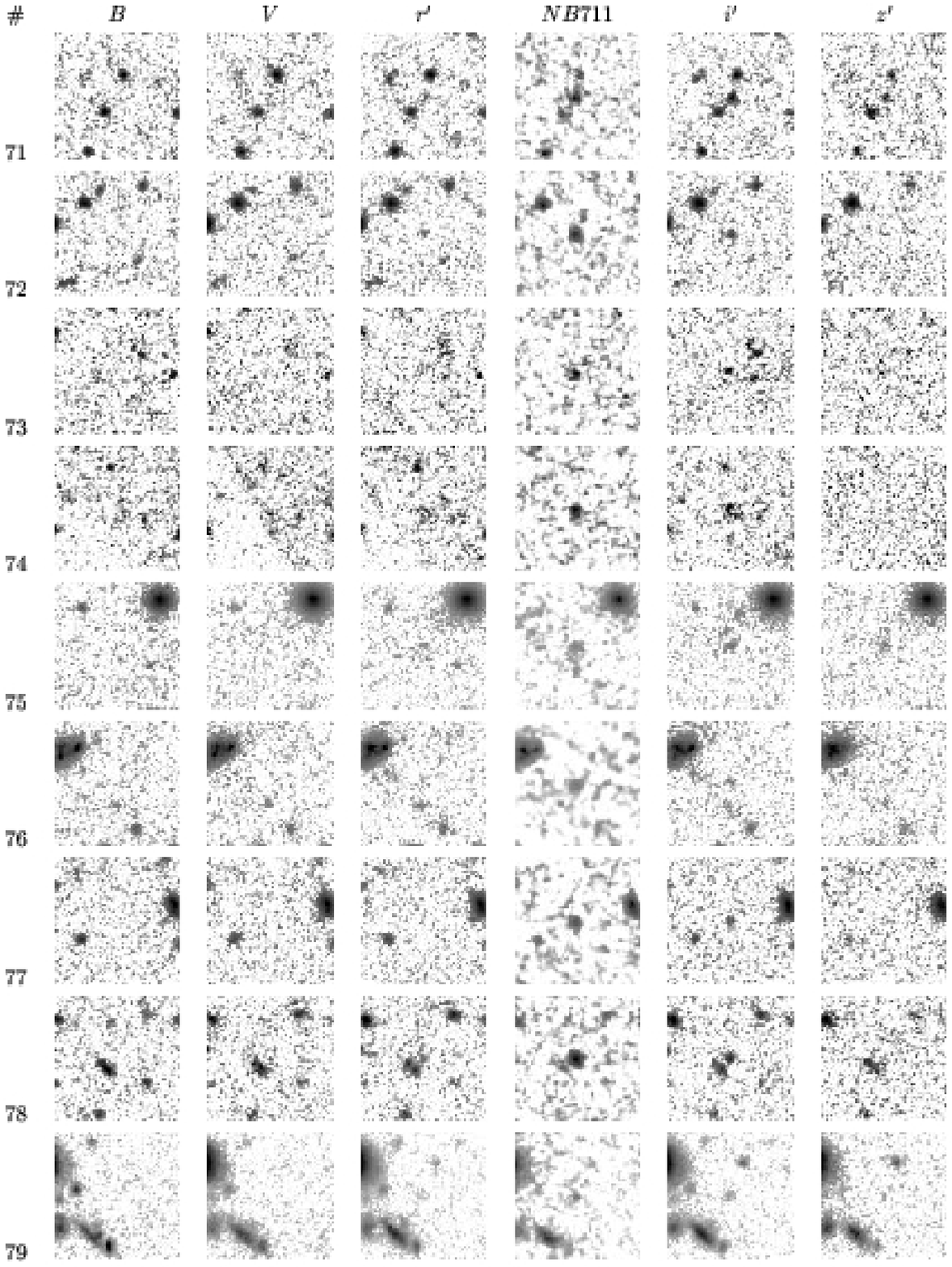}}\\[2mm]
\centerline{Fig. 4h. --- continued.}

\begin{figure}
\epsscale{1.0}
\plotone{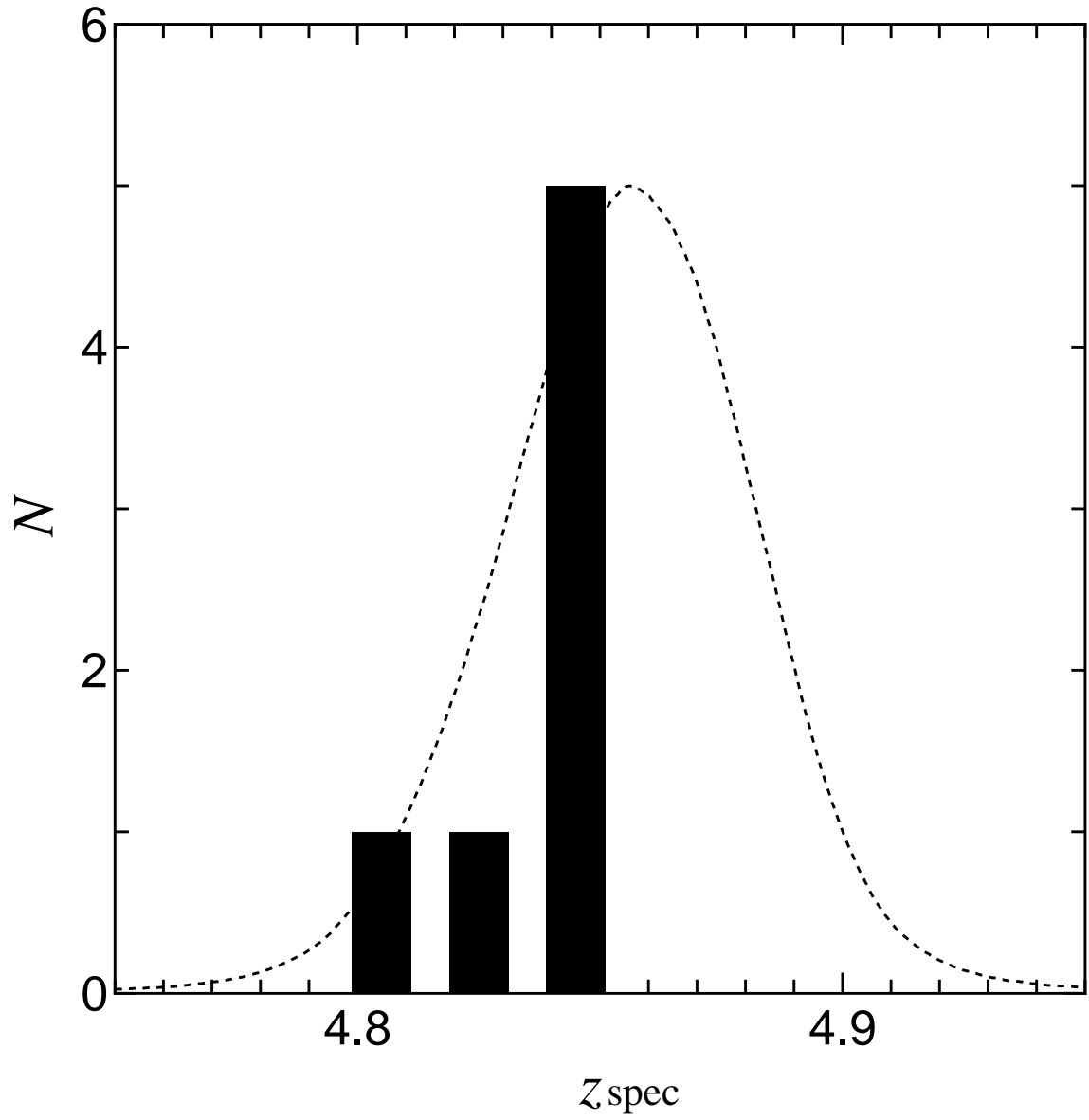}
\caption{
Spectroscopic redshift distribution of our LAE sample (7 LAEs). 
The dotted line shows the response function of the {\it NB711} band. 
\label{La:zspecdf}
}
\end{figure}

\begin{figure}
\epsscale{0.7}
\plotone{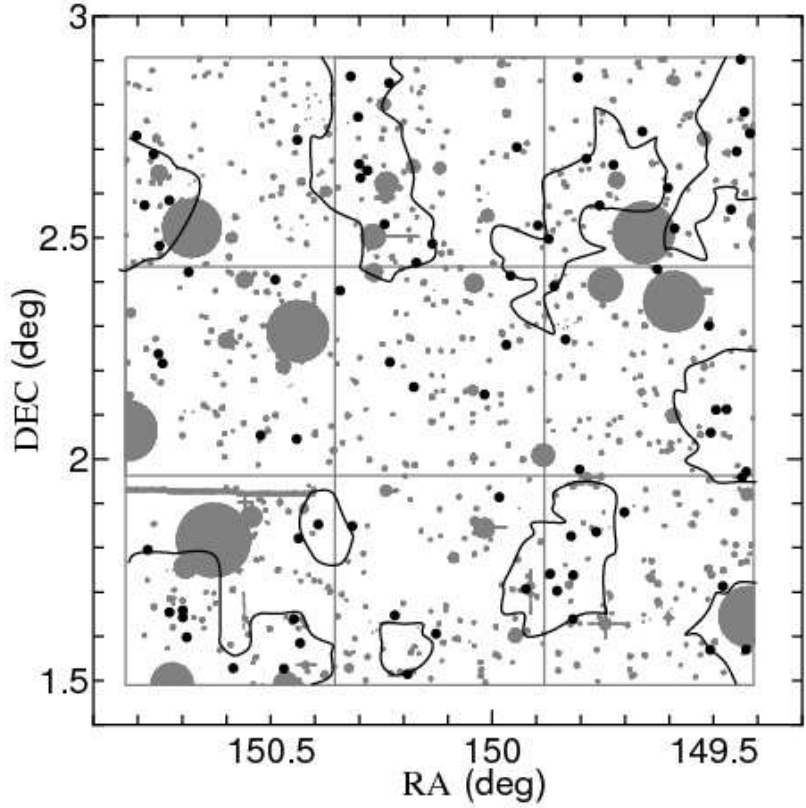}
\caption{
Spatial distributions of LAEs (black filled circles). 
The shaded regions show the areas masked out for the detection.
The contours show the local surface density of the LAEs, 
drawn at the level twice as high as the average over the field, $\rm 43 \; deg^{-2}$. 
\label{La:radec}
}
\end{figure}

\begin{figure}
\epsscale{0.8}
\plottwo{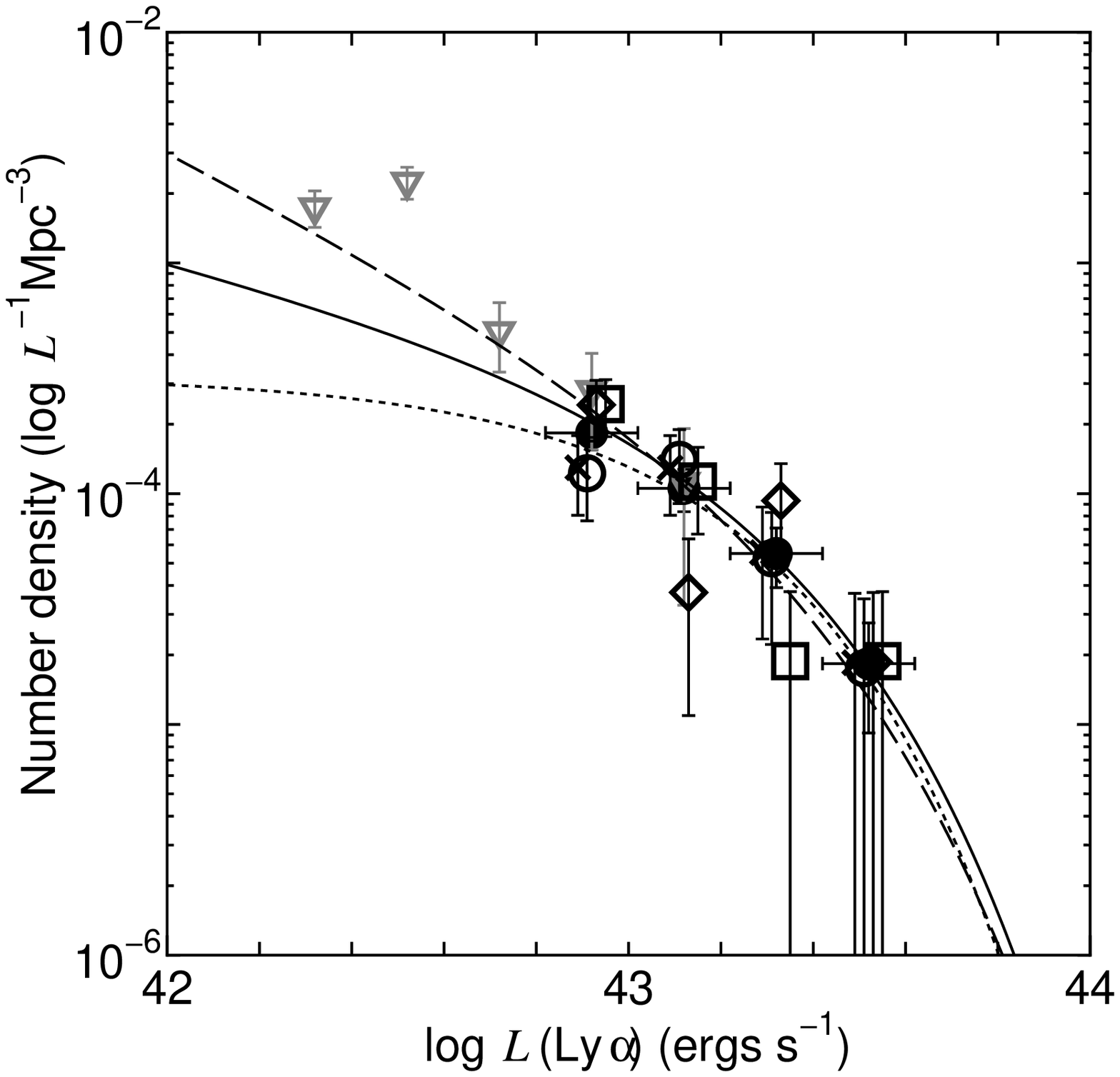}{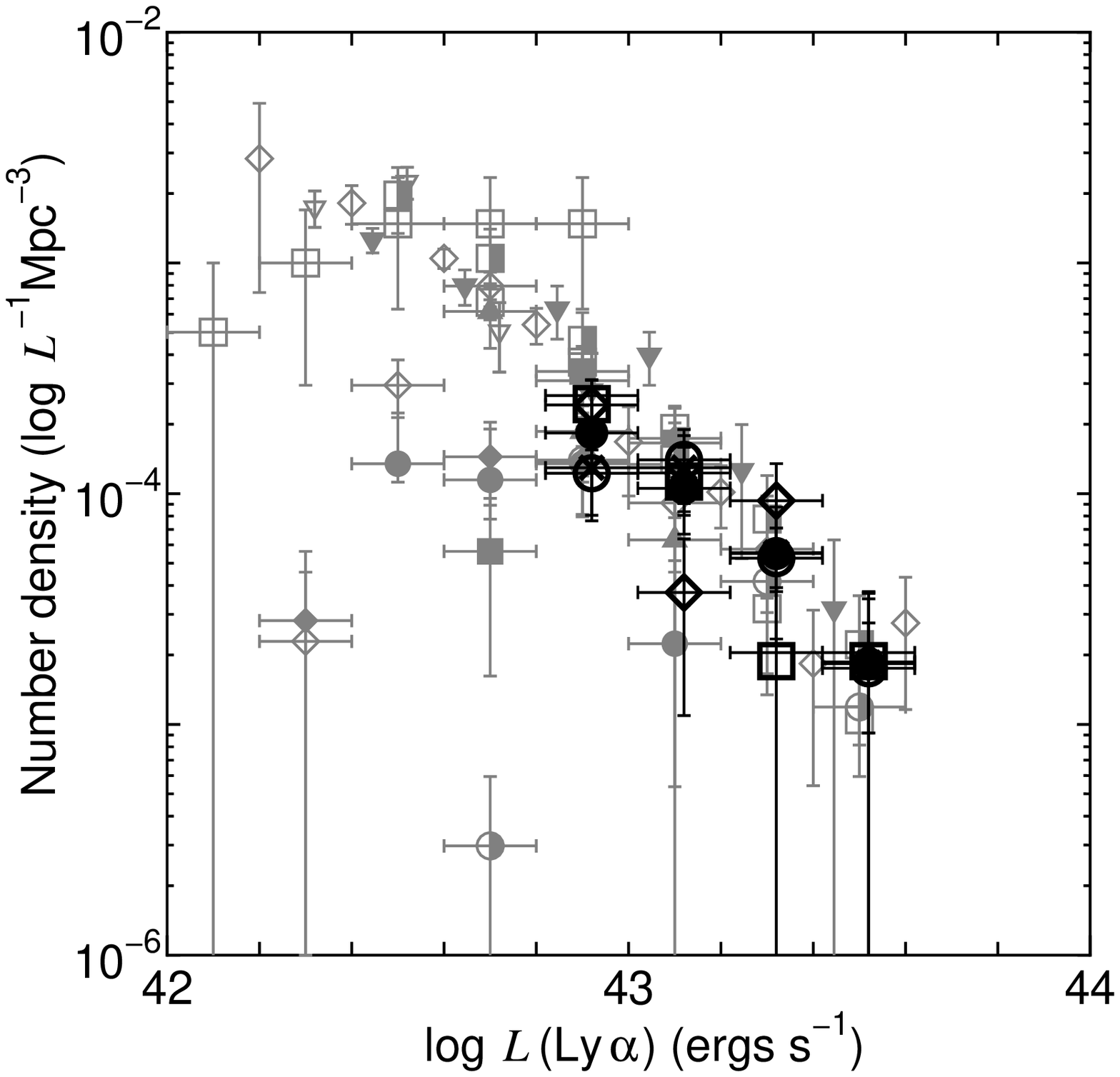}
\caption{
Left: The Ly$\alpha$ LF of our LAE sample (black symbols). 
The Ly$\alpha$ LF for the whole sample is shown with filled circles. 
The dotted, solid and dashed lines show the best-fit Schechter functions 
for the whole sample for $\alpha=-1$, $-1.5$ and $-2$, respectively. 
The Ly$\alpha$ LFs for different quadrants are shown with boxes, diamonds, 
circles, and crosses for the NE, NW, SW, and SE subfield, respectively. 
For comparison, the Ly$\alpha$ LF derived by Ouchi et al. (2003) 
is shown with inverse triangles. 
Right: Same as the left panel, compared with other surveys (gray symbols): 
for LAEs at $z \sim 3.1$ (Ouchi et al. 2008 = circles), 
$z \sim 3.4$ (Cowie \& Hu 1998 = boxes), 
$z \sim 3.7$ (Ouchi et al. 2008 = triangles), 
$z \sim 4.9$ (Ouchi et al. 2003 = inverse triangles), 
$z \sim 5.7$ (Rhoads \& Malhotra 2001 = filled circles; 
Ajiki et al. 2003 = filled boxes; 
Hu et al. 2004 = filled diamonds; 
Ajiki et al. 2006 = filled triangles; 
Shimasaku et al. 2006 = filled inverse triangles; 
Murayama et al. 2007 = half-filled circles; 
Ouchi et al. 2008 = half-filled boxes), and 
$z \sim 6.6$ (Taniguchi et al. 2005 = diamonds). 
\label{La:LaLF}
}
\end{figure}

\begin{figure}
\epsscale{0.8}
\plotone{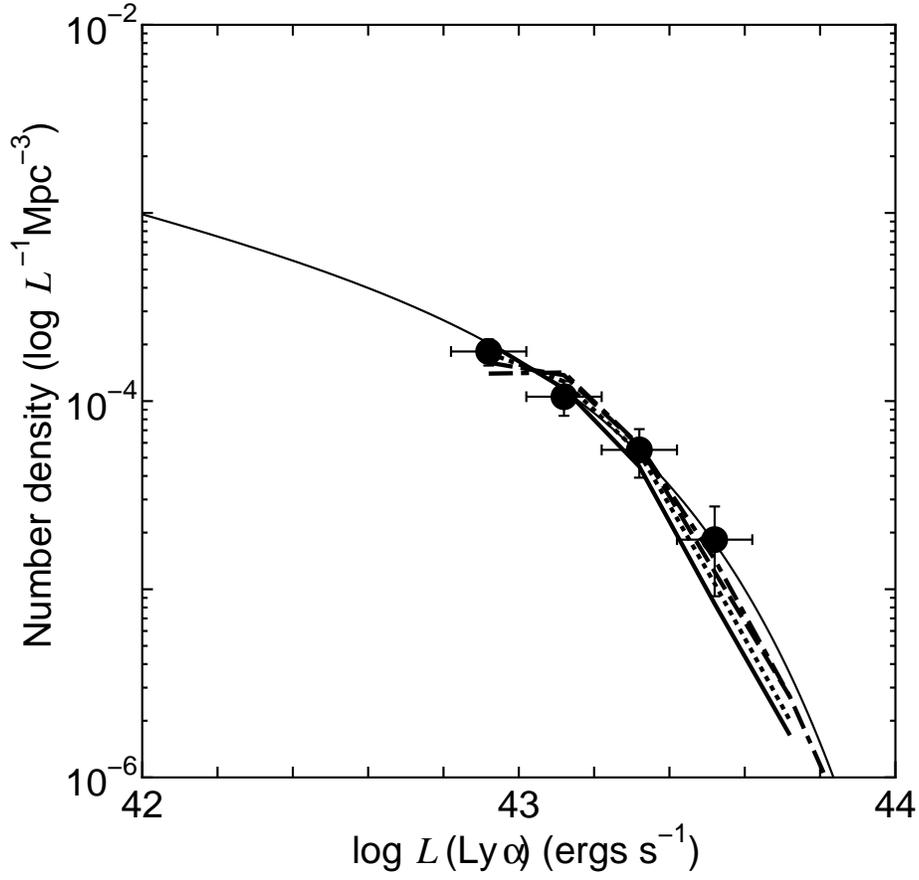}
\caption{
Results of our Monte Carlo simulations for $\alpha=-1.5$. 
The derived Ly$\alpha$ luminosity functions are 
shown as thick solid line, thick dotted line, thick dashed line, thick dash-dotted line 
for the case of $\sigma_{EW}=50$, 100, 200, \& 400 \AA, respectively. 
These luminosity functions are similar to the input Schechter function (thin solid line). 
\label{La:LaLFmock}
}
\end{figure}

\begin{figure}
\epsscale{0.5}
\plotone{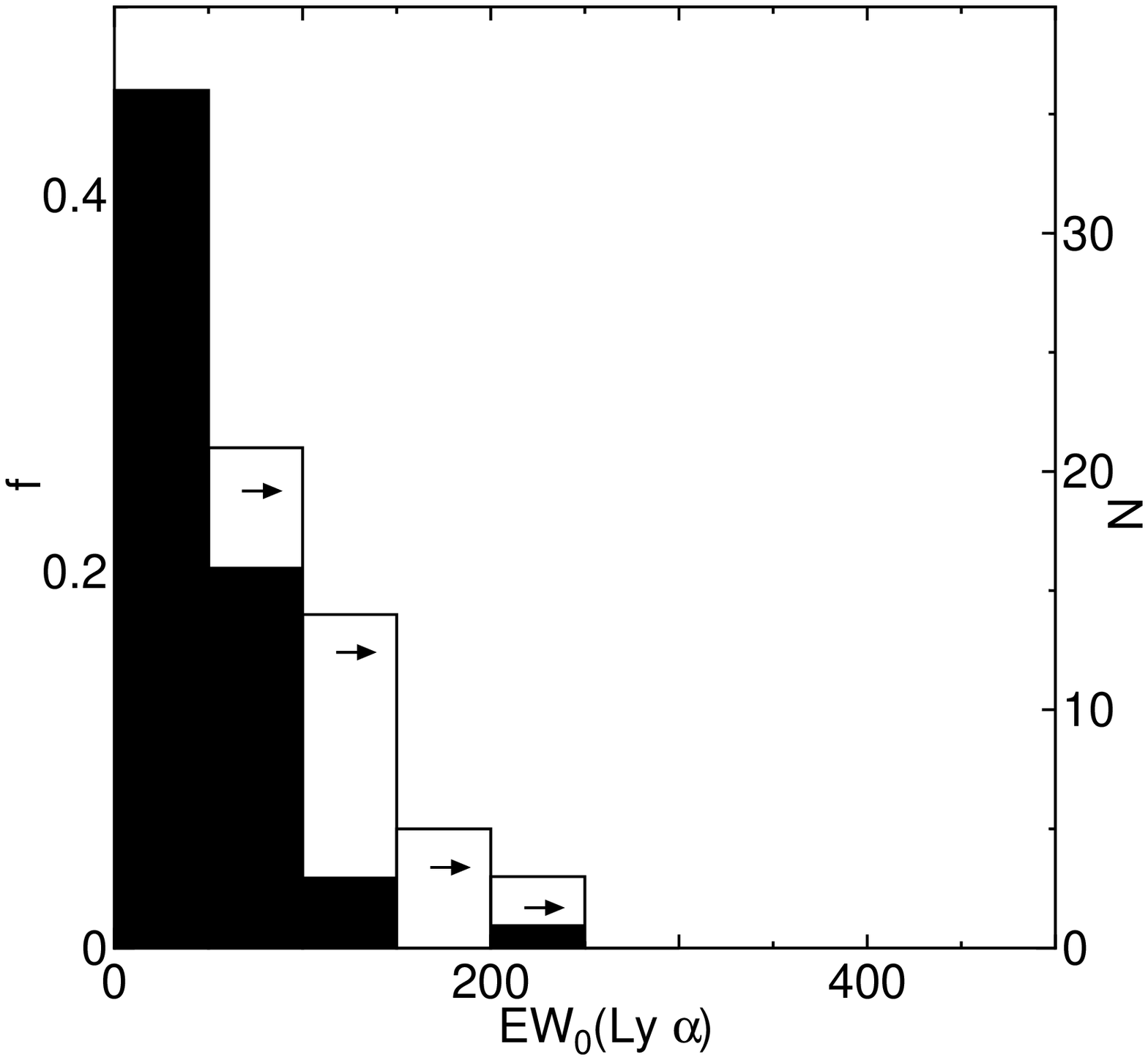}
\plotone{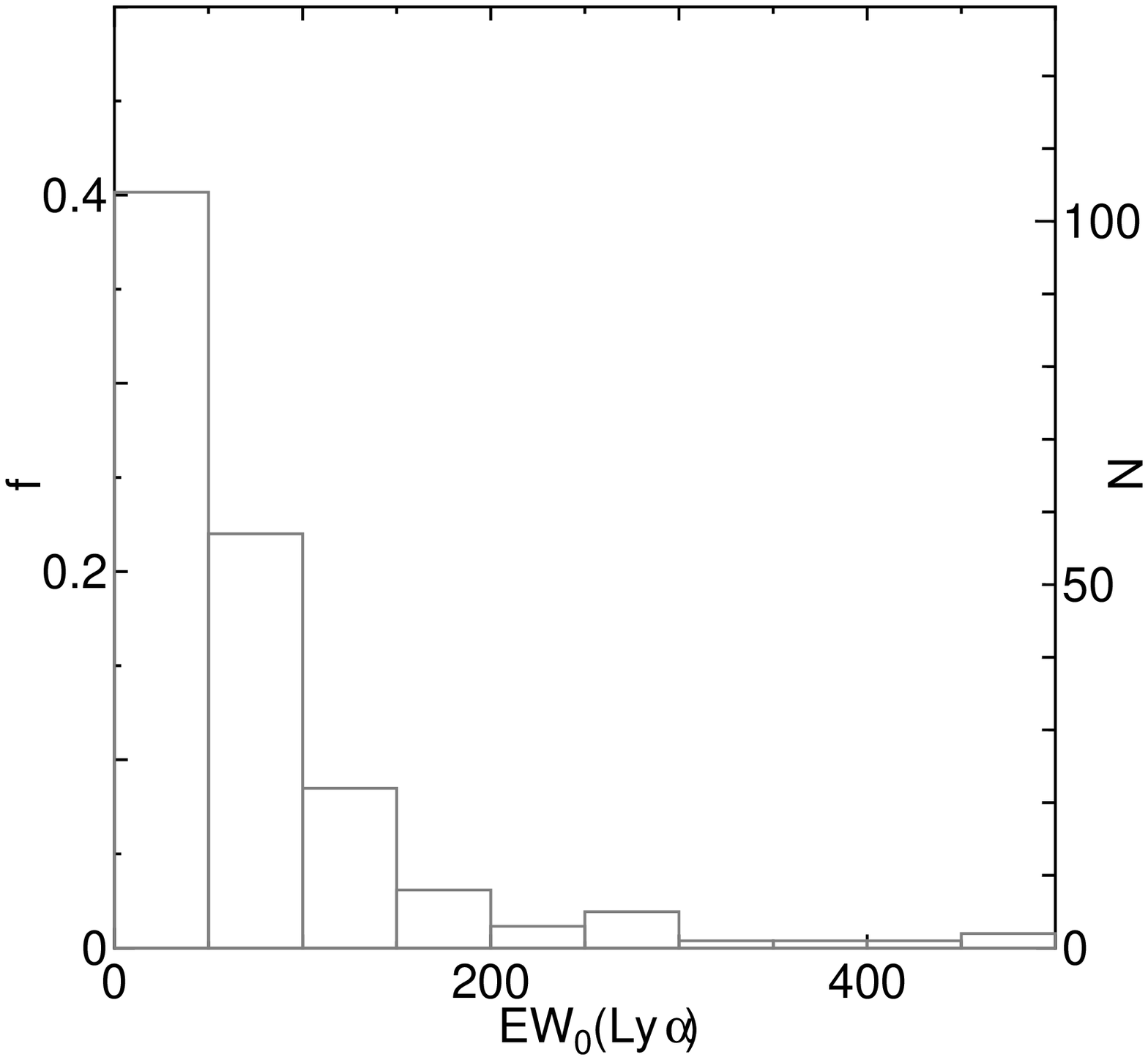}
\caption{
Top: Distribution of the rest-frame Ly$\alpha$ equivalent widths. 
Filled bars show the LAEs with the continuum detected above $1\sigma$. 
Open bars show the LAEs with no continuum detection.  
Bottom: Distribution of the rest-frame Ly$\alpha$ equivalent widths of 
the LAEs at $z \sim 3.1$ obtained by Gronwall et al. (2008).
\label{La:ewdf}
}
\end{figure}

\begin{figure}
\epsscale{1.0}
\plotone{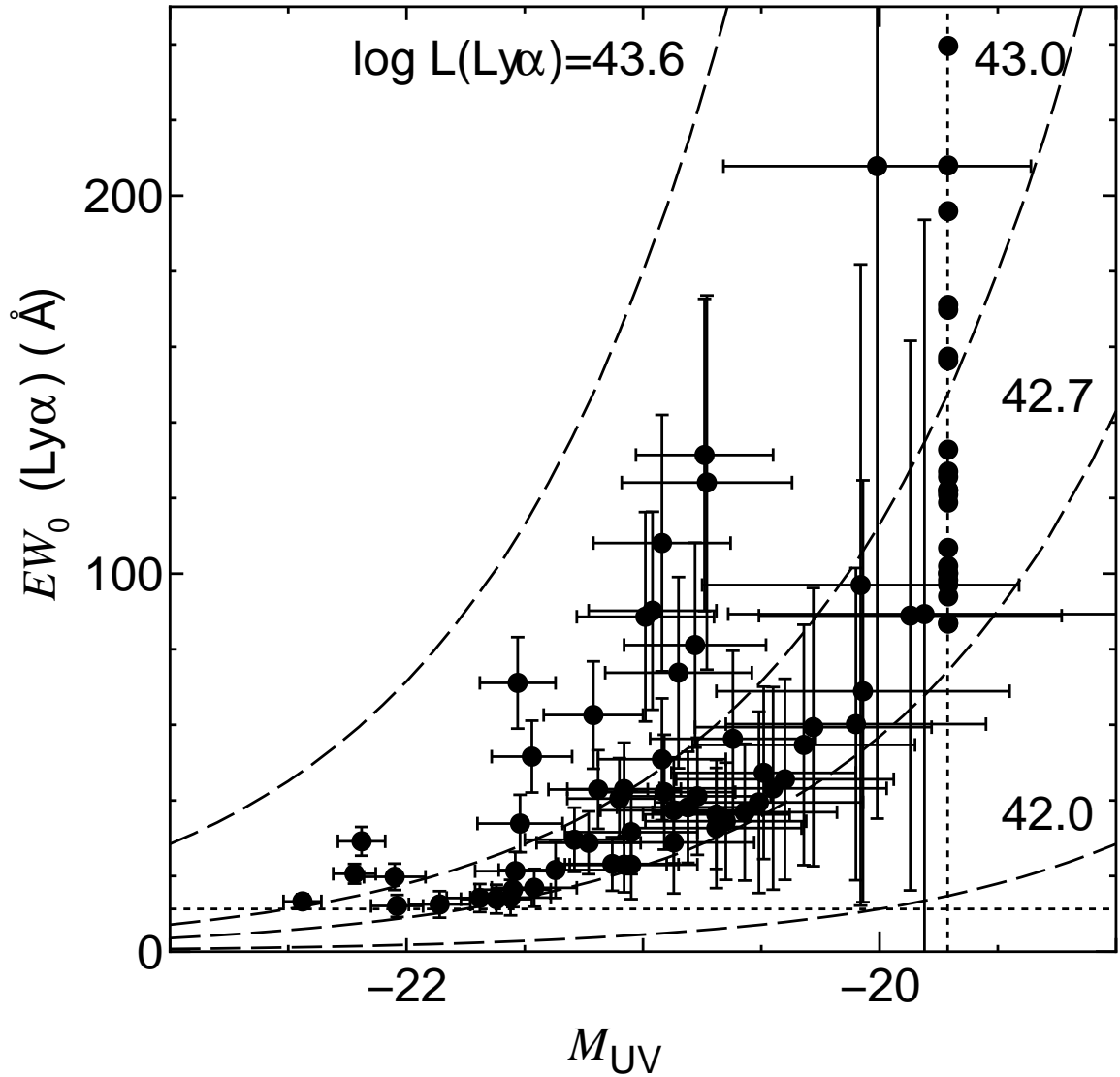}
\caption{
Rest-frame EWs of Ly$\alpha$ lines vs. absolute magnitude at 
rest-frame 1540 \AA~ for our sample of LAEs at $z\sim 4.9$. 
Dashed lines show loci of the constant Ly$\alpha$ luminosities 
for $\log L({\rm Ly}\alpha)=43.6$, 43.0 and 42.0,  
where $L({\rm Ly}\alpha)$ is in units of $\rm ergs \; s^{-1}$. 
Dotted line corresponds to $M_{\rm UV} = -19.71$ which is the rest-UV absolute magnitude 
corresponding to the $z^\prime$-band limiting magnitude ($1 \sigma$). 
\label{La:ewMuv}
}
\end{figure}

\begin{figure}
\epsscale{0.7}
\plotone{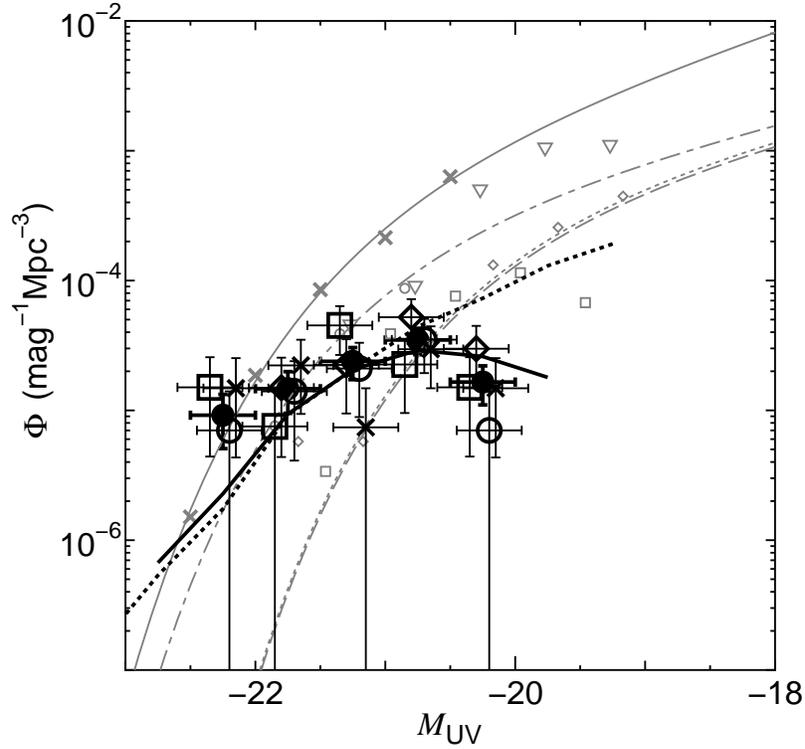}
\caption{
The rest-frame UV LF of our LAE sample (black symbols). 
The UV LF for our whole sample is shown with filled circles. 
The UV LFs for different quadrants of the COSMOS field are shown with 
black boxes, black diamonds, black circles, and black crosses for the NE, NW, SW, and SE field, respectively. 
The results of our Monte Carlo simulation for $\alpha=-1.5$ and $\sigma_{EW}=100$\AA 
are overlaid: a dotted line show the input UV LF with $EW_0({\rm Ly}\alpha)>13$\AA 
and a solid line show the output UV LF. 
For comparison, we show UV LFs from the previous surveys (gray symbols): 
LAEs at $z \sim 4.9$ (inverse triangles: Ouchi et al. 2003), 
LBGs at $z \sim 5$ (crosses and solid line: Yoshida et al. 2006), 
LAEs at $z=3.1$ (open diamonds and dotted line: Ouchi et al. 2008), 
$z=3.7$ (open boxes and dashed line: Ouchi et al. 2008), and 
$z=5.7$ (open circle and dot-dash line: Ouchi et al. 2008). 
\label{La:UVLF}
}
\end{figure}

\begin{figure}
\epsscale{0.7}
\plotone{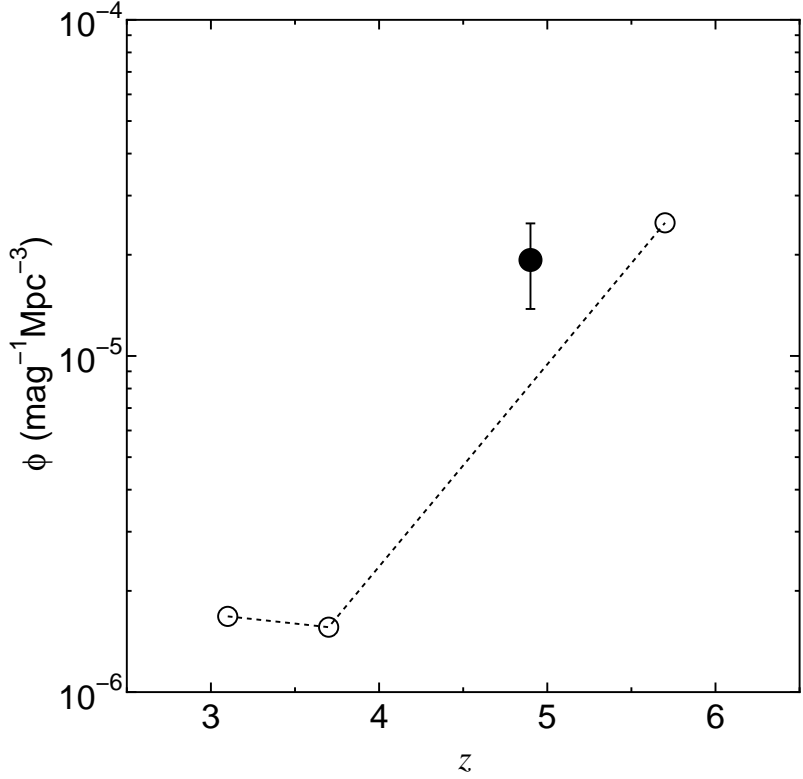}
\caption{
Number density of LAEs at $M_{\rm UV}=-21.5$ as a function of $z$. 
Our data point is shown with filled circles with a error bar. 
Open circles show the number densities derived by Ouchi et al. (2008). 
\label{La:UVLF2}
}
\end{figure}

\begin{figure}
\epsscale{1.0}
\plottwo{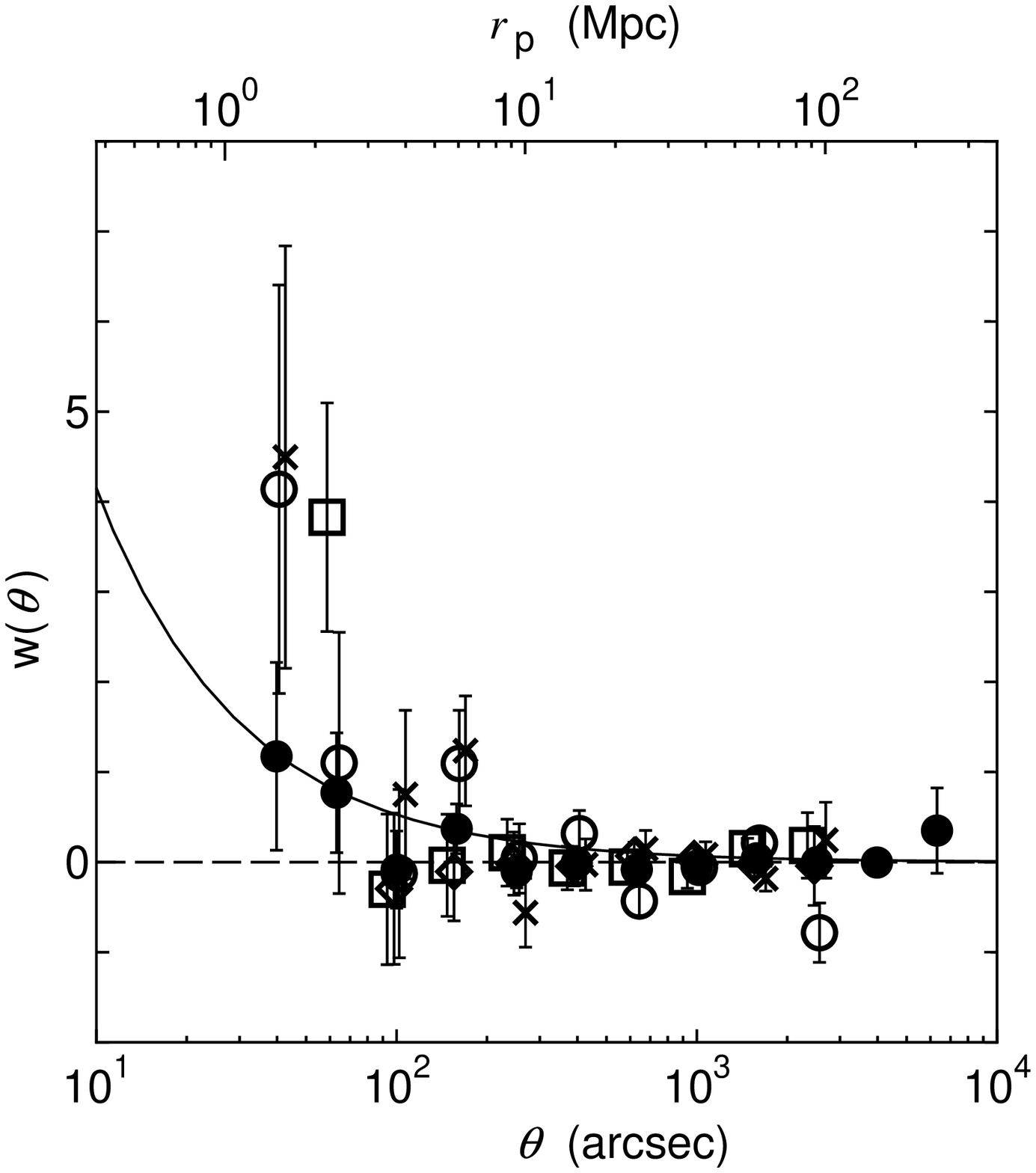}{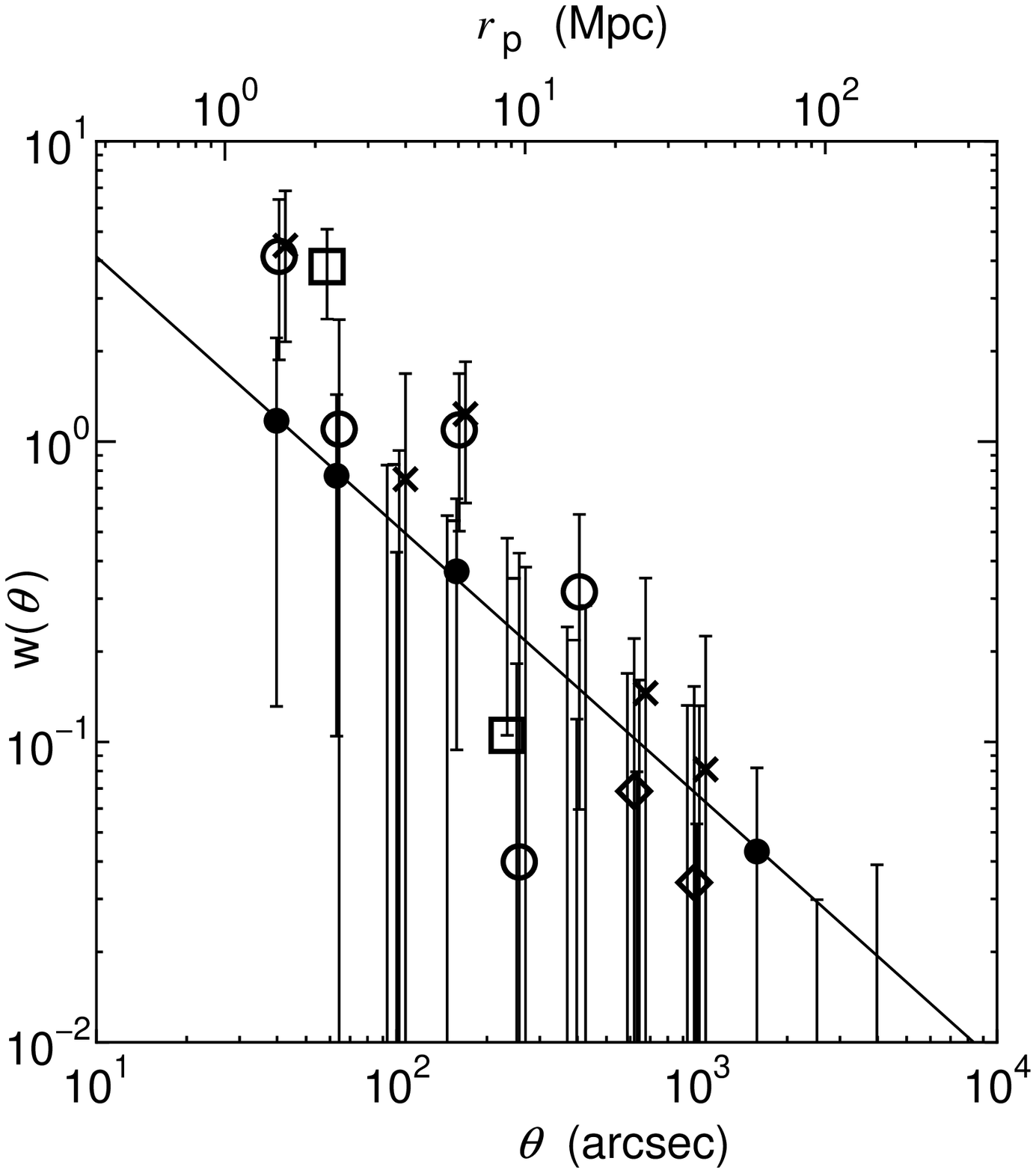}
\caption{
Left: Angular two-point correlation function (ACF) of our LAE sample. 
Filled circles show the ACF for the whole sample. 
The ACF for different quadrants are shown with boxes, diamonds, circles, and crossed 
for the NE, NW, SW and SE subfield, respectively. 
Right: Same as the left panel with $w(\theta)$ shown in logarithmic scale. 
\label{La:acf}
}
\end{figure}

\end{document}